\documentclass[11pt]{article}

\usepackage[a4paper,margin=30mm]{geometry}
\usepackage[round,authoryear]{natbib}
\usepackage{microtype}
\usepackage{pp-manuscript}
\usepackage[hidelinks]{hyperref}
\usepackage[capitalise,nameinlink,noabbrev]{cleveref}

\newcommand{\papertitle}{Systematic covariance envelopes from Wasserstein geometry:
evidence from language-model representations}

\newcommand{\paperauthors}{Marcus Gawronsky, Chun-Sung Huang}
\newcommand{\paperaffiliations}{Department of Finance and Tax, University of Cape
Town}

\newcommand{\paperdate}{August 2026}
\newcommand{\paperkeywords}{Distributional Fields; Systematic Covariance;
Wasserstein Distance; Factor Models; Language-Model Representations}
\newcommand{\paperjel}{G11; G12; C58; C60}

\ppDeclareValue{return-panel-n-observations}{1207}
\ppDeclareValue{return-panel-n-tickers}{100}
\ppDeclareValue{return-panel-date-start}{2018-03-19}
\ppDeclareValue{return-panel-date-end}{2022-12-30}

\ppDeclareValue{oos-return-panel-n-observations}{814}
\ppDeclareValue{oos-return-panel-n-tickers}{100}
\ppDeclareValue{oos-return-panel-date-start}{2023-01-04}
\ppDeclareValue{oos-return-panel-date-end}{2026-04-06}

\ppDeclareValue{n-samples-cap}{445}
\ppDeclareValue{annual-raw-article-floor}{47}
\ppDeclareValue{n-firms-news}{100}
\ppDeclareValue{matryoshka-dim-low}{64}
\ppDeclareValue{matryoshka-dim-mid}{256}
\ppDeclareValue{embed-dim}{4096}
\ppDeclareValue{sample-year-start}{2018}
\ppDeclareValue{sample-year-end}{2022}
\ppDeclareValue{text-controls-max-features}{4096}

\ppDeclareValue{corpus-raw-records}{397041}
\ppDeclareValue{corpus-articles-total}{214480}
\ppDeclareValue{corpus-duplicates-dropped}{182561}
\ppDeclareValue{corpus-year-min}{2009}
\ppDeclareValue{corpus-year-max}{2023}
\ppDeclareValue{corpus-articles-window}{129568}

\ppDeclareValue{n-firms}{100}
\ppDeclareValue{n-pairs}{4950}
\ppDeclareValue{confound-w2-coef}{1.873854}
\ppDeclareValue{confound-w2-se}{0.187977}
\ppDeclareValue{confound-w2-p}{0.001000}
\ppDeclareValue{confound-n-dyads}{4950}
\ppDeclareValue{confound-w2-ci-low}{1.515685}
\ppDeclareValue{confound-w2-ci-high}{2.250180}
\ppDeclareValue{confound-overall-r2}{0.742106}
\ppDeclareValue{confound-within-r2}{0.332462}
\ppDeclareValue{confound-w2-effect-one-sd}{0.068656}
\ppDeclareValue{confound-bootstrap-valid}{1999}

\ppDeclareValue{confound-w2-within-r2}{0.332462}
\ppDeclareValue{confound-w2-l1-tau0-coverage}{0.706667}
\ppDeclareValue{confound-w2-l1-tau0-violation}{0.293333}
\ppDeclareValue{confound-w2-l1-tau0-unresolved}{0.000000}
\ppDeclareValue{confound-w2-l1-tau0p05-coverage}{0.349091}
\ppDeclareValue{confound-w2-l1-tau0p05-violation}{0.090909}
\ppDeclareValue{confound-w2-l1-tau0p05-unresolved}{0.560000}
\ppDeclareValue{confound-w2-l1-tau0p1-coverage}{0.075960}
\ppDeclareValue{confound-w2-l1-tau0p1-violation}{0.025859}
\ppDeclareValue{confound-w2-l1-tau0p1-unresolved}{0.898182}
\ppDeclareValue{confound-w2-l1p25-tau0-coverage}{0.014949}
\ppDeclareValue{confound-w2-l1p25-tau0-violation}{0.025455}
\ppDeclareValue{confound-w2-l1p25-tau0-unresolved}{0.959596}
\ppDeclareValue{confound-w2-l1p25-tau0p05-coverage}{0.000202}
\ppDeclareValue{confound-w2-l1p25-tau0p05-violation}{0.005657}
\ppDeclareValue{confound-w2-l1p25-tau0p05-unresolved}{0.994141}
\ppDeclareValue{confound-w2-l1p25-tau0p1-coverage}{0.000000}
\ppDeclareValue{confound-w2-l1p25-tau0p1-violation}{0.001414}
\ppDeclareValue{confound-w2-l1p25-tau0p1-unresolved}{0.998586}
\ppDeclareValue{confound-w2-l1p5-tau0-coverage}{0.000000}
\ppDeclareValue{confound-w2-l1p5-tau0-violation}{0.004040}
\ppDeclareValue{confound-w2-l1p5-tau0-unresolved}{0.995960}
\ppDeclareValue{confound-w2-l1p5-tau0p05-coverage}{0.000000}
\ppDeclareValue{confound-w2-l1p5-tau0p05-violation}{0.001010}
\ppDeclareValue{confound-w2-l1p5-tau0p05-unresolved}{0.998990}
\ppDeclareValue{confound-w2-l1p5-tau0p1-coverage}{0.000000}
\ppDeclareValue{confound-w2-l1p5-tau0p1-violation}{0.000000}
\ppDeclareValue{confound-w2-l1p5-tau0p1-unresolved}{1.000000}
\ppDeclareValue{confound-firm-share-w2}{0.567909}
\ppDeclareValue{confound-firm-share-return}{0.613664}
\ppDeclareValue{confound-within-correlation}{0.576595}
\ppDeclareValue{confound-between-correlation}{0.461373}
\ppDeclareValue{confound-lofo-max-se}{0.261900}
\ppDeclareValue{confound-lofo-sign-flips}{0}
\ppDeclareValue{confound-max-leverage-dyad}{EXC--XEL}
\ppDeclareValue{confound-max-leverage}{0.034375}
\ppDeclareValue{confound-leverage-threshold}{0.040808}
\ppDeclareValue{confound-max-studentized-residual}{-6.336137}
\ppDeclareValue{confound-residual-scale-stat}{0.052203}
\ppDeclareValue{confound-residual-scale-p}{0.001000}

\ppDeclareValue{confound-fwl-slope}{1.873854}
\ppDeclareValue{confound-fwl-slope-abs-error}{0.000000}
\ppDeclareValue{confound-fwl-nuisance-columns}{100}
\ppDeclareValue{confound-fwl-omitted-firm}{AAL}
\ppDeclareValue{confound-partial-r2}{0.332462}
\ppDeclareValue{confound-sse-firm-effects}{30.318836}
\ppDeclareValue{confound-sse-firm-effects-w2}{20.238976}

\ppDeclareValue{chord-delta-one-sd}{0.068656}
\ppDeclareValue{chord-q25}{1.027830}
\ppDeclareValue{chord-rho-q25}{0.471783}
\ppDeclareValue{chord-rho-after-q25}{0.398859}
\ppDeclareValue{chord-delta-rho-q25}{-0.072924}
\ppDeclareValue{chord-median}{1.109387}
\ppDeclareValue{chord-rho-median}{0.384630}
\ppDeclareValue{chord-rho-after-median}{0.306107}
\ppDeclareValue{chord-delta-rho-median}{-0.078523}
\ppDeclareValue{chord-q75}{1.184041}
\ppDeclareValue{chord-rho-q75}{0.299024}
\ppDeclareValue{chord-rho-after-q75}{0.215375}
\ppDeclareValue{chord-delta-rho-q75}{-0.083649}
\ppDeclareValue{classification-sector-coefficient}{1.721295}
\ppDeclareValue{classification-sector-se}{0.204990}
\ppDeclareValue{classification-sector-ci-low}{1.337921}
\ppDeclareValue{classification-sector-ci-high}{2.128913}
\ppDeclareValue{classification-sector-pvalue}{0.001000}
\ppDeclareValue{classification-sector-effect-one-sd}{0.063067}
\ppDeclareValue{classification-sector-delta-correlation}{-0.071954}
\ppDeclareValue{classification-sector-anchor-correlation}{0.384630}
\ppDeclareValue{classification-sector-partial-r2}{0.227419}
\ppDeclareValue{classification-sector-n-firms}{100}
\ppDeclareValue{classification-sector-n-dyads}{4950}
\ppDeclareValue{classification-sector-bootstrap-valid-draws}{1999}
\ppDeclareValue{classification-industry-coefficient}{1.554714}
\ppDeclareValue{classification-industry-se}{0.205098}
\ppDeclareValue{classification-industry-ci-low}{1.166991}
\ppDeclareValue{classification-industry-ci-high}{1.967252}
\ppDeclareValue{classification-industry-pvalue}{0.001000}
\ppDeclareValue{classification-industry-effect-one-sd}{0.056963}
\ppDeclareValue{classification-industry-delta-correlation}{-0.064817}
\ppDeclareValue{classification-industry-anchor-correlation}{0.384630}
\ppDeclareValue{classification-industry-partial-r2}{0.184065}
\ppDeclareValue{classification-industry-n-firms}{100}
\ppDeclareValue{classification-industry-n-dyads}{4950}
\ppDeclareValue{classification-industry-bootstrap-valid-draws}{1999}
\ppDeclareValue{classification-within-industry-coefficient}{1.444067}
\ppDeclareValue{classification-within-industry-se}{0.494869}
\ppDeclareValue{classification-within-industry-ci-low}{0.530278}
\ppDeclareValue{classification-within-industry-ci-high}{2.441902}
\ppDeclareValue{classification-within-industry-pvalue}{0.013000}
\ppDeclareValue{classification-within-industry-effect-one-sd}{0.054247}
\ppDeclareValue{classification-within-industry-delta-correlation}{-0.049728}
\ppDeclareValue{classification-within-industry-anchor-correlation}{0.604331}
\ppDeclareValue{classification-within-industry-partial-r2}{0.309414}
\ppDeclareValue{classification-within-industry-n-firms}{86}
\ppDeclareValue{classification-within-industry-n-dyads}{182}
\ppDeclareValue{classification-within-industry-bootstrap-valid-draws}{1999}
\ppDeclareValue{classification-etnic-roster-coefficient}{1.765123}
\ppDeclareValue{classification-etnic-roster-se}{0.252055}
\ppDeclareValue{classification-etnic-roster-ci-low}{1.297325}
\ppDeclareValue{classification-etnic-roster-ci-high}{2.279645}
\ppDeclareValue{classification-etnic-roster-pvalue}{0.001000}
\ppDeclareValue{classification-etnic-roster-effect-one-sd}{0.062758}
\ppDeclareValue{classification-etnic-roster-delta-correlation}{-0.070297}
\ppDeclareValue{classification-etnic-roster-anchor-correlation}{0.407320}
\ppDeclareValue{classification-etnic-roster-partial-r2}{0.227563}
\ppDeclareValue{classification-etnic-roster-n-firms}{81}
\ppDeclareValue{classification-etnic-roster-n-dyads}{3240}
\ppDeclareValue{classification-etnic-roster-bootstrap-valid-draws}{1999}
\ppDeclareValue{classification-etnic-coefficient}{1.313738}
\ppDeclareValue{classification-etnic-se}{0.258404}
\ppDeclareValue{classification-etnic-ci-low}{0.857583}
\ppDeclareValue{classification-etnic-ci-high}{1.880262}
\ppDeclareValue{classification-etnic-pvalue}{0.001000}
\ppDeclareValue{classification-etnic-effect-one-sd}{0.046709}
\ppDeclareValue{classification-etnic-delta-correlation}{-0.051945}
\ppDeclareValue{classification-etnic-anchor-correlation}{0.407320}
\ppDeclareValue{classification-etnic-partial-r2}{0.112393}
\ppDeclareValue{classification-etnic-n-firms}{81}
\ppDeclareValue{classification-etnic-n-dyads}{3240}
\ppDeclareValue{classification-etnic-bootstrap-valid-draws}{1999}
\ppDeclareValue{classification-etnic-control-coefficient}{-0.233846}
\ppDeclareValue{classification-etnic-control-se}{0.054750}
\ppDeclareValue{classification-etnic-control-ci-low}{-0.343407}
\ppDeclareValue{classification-etnic-control-ci-high}{-0.128983}
\ppDeclareValue{classification-etnic-control-effect-one-sd}{-0.022161}
\ppDeclareValue{classification-etnic-control-partial-r2}{0.053310}
\ppDeclareValue{classification-etnic-control-correlation-with-w2}{-0.546430}
\ppDeclareValue{classification-n-industries}{38}
\ppDeclareValue{etnic-n-firms}{81}
\ppDeclareValue{etnic-year-start}{2018}
\ppDeclareValue{etnic-year-end}{2022}
\ppDeclareValue{etnic-matrix-correlation-w2}{0.546430}
\ppDeclareValue{etnic-matrix-correlation-w2-spearman}{0.396029}
\ppDeclareValue{etnic-matrix-correlation-w2-partial-sector}{0.326985}
\ppDeclareValue{etnic-matrix-correlation-pvalue-max}{0.000100}
\ppDeclareValue{etnic-matrix-correlation-permutations}{9999}

\ppDeclareValue{w2-permanova-n-firms}{100}
\ppDeclareValue{w2-permanova-n-groups}{9}
\ppDeclareValue{w2-permanova-r2}{0.136479}
\ppDeclareValue{w2-permanova-f}{1.797814}
\ppDeclareValue{w2-permanova-p}{0.000100}
\ppDeclareValue{w2-permanova-null-r2}{0.080808}
\ppDeclareValue{w2-permanova-excess-r2}{0.055676}
\ppDeclareValue{w2-permanova-dispersion-p}{0.000100}

\ppDeclareValue{sample-desc-articles-per-firm}{128.000000}
\ppDeclareValue{sample-desc-total-articles}{12800}

\ppDeclareValue{matryoshka-preserve-64-spearman}{0.803636}
\ppDeclareValue{matryoshka-preserve-256-spearman}{0.925257}

\ppDeclareValue{pcoa-stress-1}{0.776467}
\ppDeclareValue{pcoa-shepard-pearson}{0.546199}
\ppDeclareValue{pcoa-shepard-spearman}{0.514484}
\ppDeclareValue{pcoa-top5-neighbour-recall}{0.252000}
\ppDeclareValue{pcoa-exact-nearest-preserved}{14}
\ppDeclareValue{pcoa-two-axis-share}{0.064480}
\ppDeclareValue{pcoa-negative-inertia-share}{0.000000}

\ppDeclareValue{metric-mds-stress-1}{0.396103}
\ppDeclareValue{metric-mds-shepard-pearson}{0.553424}
\ppDeclareValue{metric-mds-shepard-spearman}{0.553463}
\ppDeclareValue{metric-mds-top5-neighbour-recall}{0.270000}
\ppDeclareValue{metric-mds-exact-nearest-preserved}{9}

\ppDeclareValue{neighbour-holdout-n-firms}{100}
\ppDeclareValue{neighbour-holdout-n-pairs}{4950}
\ppDeclareValue{neighbour-holdout-n-selected}{7}
\ppDeclareValue{neighbour-holdout-n-dates}{503}
\ppDeclareValue{neighbour-holdout-min-prearticles}{32}
\ppDeclareValue{neighbour-holdout-training-year-start}{2018}
\ppDeclareValue{neighbour-holdout-training-year-end}{2020}
\ppDeclareValue{neighbour-holdout-pool-year-end}{2022}
\ppDeclareValue{neighbour-holdout-evaluation-year-start}{2021}
\ppDeclareValue{neighbour-holdout-evaluation-year-end}{2022}
\ppDeclareValue{neighbour-holdout-communication-services-w2}{0.965}
\ppDeclareValue{neighbour-holdout-communication-services-rms-gap}{0.302}
\ppDeclareValue{neighbour-holdout-communication-services-terminal-gap}{0.540}
\ppDeclareValue{neighbour-holdout-communication-services-percentile}{38.9}
\ppDeclareValue{neighbour-holdout-consumer-cyclical-w2}{0.968}
\ppDeclareValue{neighbour-holdout-consumer-cyclical-rms-gap}{0.239}
\ppDeclareValue{neighbour-holdout-consumer-cyclical-terminal-gap}{0.057}
\ppDeclareValue{neighbour-holdout-consumer-cyclical-percentile}{29.6}
\ppDeclareValue{neighbour-holdout-consumer-defensive-w2}{0.995}
\ppDeclareValue{neighbour-holdout-consumer-defensive-rms-gap}{0.067}
\ppDeclareValue{neighbour-holdout-consumer-defensive-terminal-gap}{0.104}
\ppDeclareValue{neighbour-holdout-consumer-defensive-percentile}{0.0}
\ppDeclareValue{neighbour-holdout-healthcare-w2}{0.983}
\ppDeclareValue{neighbour-holdout-healthcare-rms-gap}{0.208}
\ppDeclareValue{neighbour-holdout-healthcare-terminal-gap}{0.206}
\ppDeclareValue{neighbour-holdout-healthcare-percentile}{25.6}
\ppDeclareValue{neighbour-holdout-industrials-w2}{0.961}
\ppDeclareValue{neighbour-holdout-industrials-rms-gap}{0.182}
\ppDeclareValue{neighbour-holdout-industrials-terminal-gap}{0.212}
\ppDeclareValue{neighbour-holdout-industrials-percentile}{60.0}
\ppDeclareValue{neighbour-holdout-technology-w2}{0.939}
\ppDeclareValue{neighbour-holdout-technology-rms-gap}{0.392}
\ppDeclareValue{neighbour-holdout-technology-terminal-gap}{0.743}
\ppDeclareValue{neighbour-holdout-technology-percentile}{78.3}
\ppDeclareValue{neighbour-holdout-utilities-w2}{0.882}
\ppDeclareValue{neighbour-holdout-utilities-rms-gap}{0.251}
\ppDeclareValue{neighbour-holdout-utilities-terminal-gap}{0.321}
\ppDeclareValue{neighbour-holdout-utilities-percentile}{100.0}

\ppDeclareValue{oos-dyadic-n-firms}{100}
\ppDeclareValue{oos-dyadic-n-dyads}{4950}
\ppDeclareValue{oos-dyadic-design-columns}{101}
\ppDeclareValue{oos-dyadic-residual-dof}{4849}
\ppDeclareValue{oos-dyadic-text-universe-firms}{100}
\ppDeclareValue{oos-dyadic-attrition-firms}{0}
\ppDeclareValue{oos-dyadic-w2-coef}{1.945060}
\ppDeclareValue{oos-dyadic-w2-se}{0.243170}
\ppDeclareValue{oos-dyadic-w2-p}{0.001000}
\ppDeclareValue{oos-dyadic-w2-ci-low}{1.515894}
\ppDeclareValue{oos-dyadic-w2-ci-high}{2.453857}
\ppDeclareValue{oos-dyadic-w2-effect-one-sd}{0.071265}
\ppDeclareValue{oos-dyadic-w2-within-r2}{0.313726}
\ppDeclareValue{text-controls-tfidf-document-partial-r2}{0.132819}
\ppDeclareValue{source-pool-total-articles}{43720}
\ppDeclareValue{classification-identifying-industries}{7}
\ppDeclareValue{classification-identifying-firms}{45}

\ppDeclareValue{lean-verified-lem:polarization}{checked}
\ppDeclareValue{lean-disclosure-lem:polarization}{The polarization identity is provided by mathlib.}
\ppDeclareValue{lean-verified-thm:random-exposure-envelope}{checked}
\ppDeclareValue{lean-disclosure-thm:random-exposure-envelope}{Machine-checked for probability laws on a separable real Hilbert space under the stated integrability hypotheses and a supplied optimal-plan certificate. The certified plan attains the covariance ceiling.}
\ppDeclareValue{lean-verified-eq:random-exposure-gap}{checked}
\ppDeclareValue{lean-disclosure-eq:random-exposure-gap}{Machine-checked as an exact polarization identity for an integrable coupling of probability laws on a separable real Hilbert space. The covariance gap equals half the coupling's excess transport cost.}
\ppDeclareValue{lean-verified-eq:random-exposure-excess-nonneg}{checked}
\ppDeclareValue{lean-disclosure-eq:random-exposure-excess-nonneg}{Machine-checked for integrable couplings of probability laws on a separable real Hilbert space. Nonnegativity follows from evaluating excess relative to a certified minimum transport cost.}
\ppDeclareValue{lean-verified-eq:random-w2-upper-metric}{checked}
\ppDeclareValue{lean-disclosure-eq:random-w2-upper-metric}{Machine-checked upper metric bound for deterministic synchronous transmission maps and their pushforward second-moment probability laws. Conditional stochastic-channel transmission requires an additional reduction not certified by this declaration. The lower metric bound is established analytically in the accompanying proof.}
\ppDeclareValue{lean-verified-thm:random-exposure-lower-endpoint}{checked}
\ppDeclareValue{lean-disclosure-thm:random-exposure-lower-endpoint}{Machine-checked lower covariance endpoint for second-moment probability laws on a separable real Hilbert space, using reflection and a supplied optimal-plan certificate.}
\ppDeclareValue{lean-verified-eq:random-w2-upper-transfer}{checked}
\ppDeclareValue{lean-disclosure-eq:random-w2-upper-transfer}{Machine-checked upper bound on optimized squared transport costs for finite clouds, under the stated transport certificates, Lipschitz condition, slack tolerances, and support-diameter bound.}
\ppDeclareValue{lean-verified-eq:random-w2-lower-transfer}{checked}
\ppDeclareValue{lean-disclosure-eq:random-w2-lower-transfer}{Machine-checked lower bound on optimized squared transport costs for finite clouds, under the stated transport certificates, positive lower Lipschitz condition, slack tolerances, and support-diameter bound.}
\ppDeclareValue{lean-verified-eq:total-return-excess-correction}{checked}
\ppDeclareValue{lean-disclosure-eq:total-return-excess-correction}{Machine-checked for an independent isotropic factor law in risk coordinates under the stated integrability, centering, and orthogonality conditions. Total-return squared distance minus optimal exposure cost equals systematic excess plus residual variance. Standardization to correlation units is the empirical application of this identity.}

\newcommand{\ppLeanStatusList}{%
  \begin{itemize}
    \item \Cref{lem:polarization}: \ppvalue{lean-disclosure-lem:polarization}
    \item \Cref{thm:random-exposure-envelope}: \ppvalue{lean-disclosure-thm:random-exposure-envelope}
    \item \Cref{eq:random-exposure-gap}: \ppvalue{lean-disclosure-eq:random-exposure-gap}
    \item \Cref{eq:random-exposure-excess-nonneg}: \ppvalue{lean-disclosure-eq:random-exposure-excess-nonneg}
    \item \Cref{eq:random-w2-upper-metric}: \ppvalue{lean-disclosure-eq:random-w2-upper-metric}
    \item \Cref{thm:random-exposure-lower-endpoint}: \ppvalue{lean-disclosure-thm:random-exposure-lower-endpoint}
    \item \Cref{eq:random-w2-upper-transfer}: \ppvalue{lean-disclosure-eq:random-w2-upper-transfer}
    \item \Cref{eq:random-w2-lower-transfer}: \ppvalue{lean-disclosure-eq:random-w2-lower-transfer}
    \item \Cref{eq:total-return-excess-correction}: \ppvalue{lean-disclosure-eq:total-return-excess-correction}
  \end{itemize}%
}

\hypersetup{
  pdftitle={\papertitle},
  pdfauthor={\paperauthors},
  pdfsubject={Distribution-valued factor covariance model},
  pdfkeywords={\paperkeywords}
}

\title{\papertitle}
\author{\paperauthors\\\paperaffiliations}
\date{\paperdate}

\begin{document}

\maketitle

\begin{abstract}
  Firm characteristics are commonly represented as fixed vectors, even though
evidence about firms' operations arrives as heterogeneous collections of
articles and reports.
We represent firms by distributions and derive restrictions on the systematic
covariance their factor exposures can support.
Quadratic Wasserstein distance measures the minimum displacement needed to
align two distributions; in exposure space, it determines their sharp
covariance ceiling.
A common characteristic-to-exposure benchmark with bounded distortion and
firm-specific departures transfers this restriction to characteristic space.
The framework makes explicit how transmission uncertainty widens the range
of feasible covariance ceilings.
We construct firm distributions from language-model embeddings of financial
news, disclosures, and analyst reports for Nasdaq firms over
\ppvalue{sample-year-start}--\ppvalue{sample-year-end}.
Greater article-embedding distance is associated with weaker return
co-movement after symmetric firm effects: the return-distance coefficient is
\ppnum[3]{confound-w2-coef}, with 95\% interval
[\ppnum[3]{confound-w2-ci-low}, \ppnum[3]{confound-w2-ci-high}].
The association remains positive after sector, industry, and 10-K
product-market controls, and retains its sign in a later return window.
A sensitivity grid shows how quickly the covariance restriction loses
informativeness as transmission tolerances widen.
Together, the framework and evidence support screening firm pairs for shared
exposures and assessing covariance under explicit economic scenarios.

\end{abstract}


\medskip

\noindent\textit{JEL classification:} \paperjel

\section{Introduction}\label{sec:introduction}

Factor models represent systematic return variation through exposures to
common shocks, and the resulting covariances determine how firm risks
aggregate in a portfolio
\citep{ross_arbitrage_1976,fama_multifactor_1996,daniel_evidence_1997}.
The recent wave of large initial public offerings makes this dependence
question especially consequential: a prospective holding can have substantial
economic importance and little public trading history, alongside years of
reporting about its products, customers, suppliers, and contracts.
Nasdaq's record first-half listings in 2026 illustrate the scale of this
information mismatch \citep{nasdaq_listings_2026}.
An investment manager needs to understand how those activities relate to risks
already present in the portfolio, including connections obscured by sector
labels or diversified business models.

Language models make this documentary evidence increasingly accessible.
\citet{lopezlira_chatgpt_2026} document that headline assessments capture
initial stock-price responses and predict subsequent drift.
Assessing an individual company's news raises a further portfolio question:
how do its risks fit together with those of other holdings?
The characteristics used to describe factor exposures are usually encoded as
points or finite vectors.
Yet evidence about a firm arrives as a collection of reports about its
products, costs, contracts, and markets.
Retaining that collection as a distribution preserves heterogeneity within
the firm and raises a financial question: what does separation between two
such distributions imply for the systematic risk the firms can share?

We derive a covariance restriction from the geometry of those distributions.
Quadratic Wasserstein distance, denoted W$_2$, measures the minimum
root-mean-square displacement needed to align their probability mass.
A coupling specifies which draws from the two distributions occur together.
In factor-exposure space, the coupling with the smallest squared displacement
attains the greatest systematic covariance.
This follows from polarization: covariance equals average exposure magnitudes
minus half the expected squared distance between exposures.
The distance between the marginal exposure laws therefore determines a sharp
covariance ceiling; the realized coupling falls below it by half its excess
transport cost.

To connect that result to firm information, we introduce a common benchmark
relationship between economic characteristics and factor exposures, allowing
bounded firm-specific departures.
The benchmark describes how activities and contracts transmit common shocks
to equity returns: fuel intensity, fare flexibility, hedging, and leverage,
for example, shape an airline's exposure to fuel prices.
Bounds on the distortion of characteristic distances and on departures from
the benchmark yield an interval for the covariance ceiling.
\Cref{fig:continuous-bound-construction} summarizes this construction.

\begin{figure}[H]
  \centering
  \import{images/}{continuous_bound_construction\ppfiguresuffix.pgf}
  \caption{Construction of the one-sided systematic-covariance envelope.
    Panel A represents each firm's characteristic law.
    Panel B maps it through the firm-specific transmission kernel $K_i$ to
    its exposure law $P_i$, with a common carrier and bounded departures.
    Panel C combines the exposure-law covariance ceiling with transfer bounds
    to form the conditional interval evaluated from observed articles.
  The operator $\Gamma$ fixes the common risk-coordinate geometry.}
  \label{fig:continuous-bound-construction}
\end{figure}

Empirically, we represent each firm by a distribution of language-model
embeddings of financial news, disclosures, and analyst reports for a Nasdaq
sample spanning \ppvalue{sample-year-start}--\ppvalue{sample-year-end}.
News production gives this representation economic content: investors demand
information about cash flows and risks, while reporting costs and audience
interest influence which activities receive coverage.
The empirical design explains this selection mechanism and relates distances
between the resulting article distributions to return co-movement.

In a dyadic regression with symmetric firm effects, one standard deviation
more article-embedding W$_2$ corresponds to roughly eight fewer correlation points.
The positive return-distance association persists after sector and finer
industry matching, within industries, and after controlling for
Hoberg--Phillips 10-K product-market similarity.
It also retains its sign when returns are measured in a later window.
A separate sensitivity grid evaluates the covariance restriction over
specified transmission scenarios.
Its informativeness declines rapidly as multiplicative distortion and
firm-specific slack increase, showing how tightly the transmission
relationship must be specified for the restriction to discriminate among
candidate covariances.

Controlled comparisons across encoders, embedding widths, distance costs,
and reported training-information vintages examine how the representation
measures this relationship.
Together with transport's allocation and adjustment-cost interpretations,
they connect the covariance argument to economically interpretable
measurement choices.

The contribution connects a distributional representation of firm information,
a sharp covariance result, and evidence on return co-movement.
For practitioners, article distance supplies a way to screen firm pairs for
shared risks across conventional classifications.
The covariance envelope then organizes an assessment of those pairs under
explicit assumptions about how their economic characteristics translate into
systematic exposures.

\Cref{sec:literature} relates the contribution to characteristic-based loading
models and textual proximity measures.
\Cref{sec:methodology} develops the economic transmission benchmark and
covariance envelope.
\Cref{sec:empirical-design,sec:results} construct the article distributions
and report the association, classification comparisons, and sensitivity evidence.

\section{Literature review}\label{sec:literature}

Three literatures connect firm information to systematic covariance:
textual analysis in finance, optimal transport, and characteristic-based and
spatial representations of systematic risk.
Together, these strands provide the measurement object, the distance used to
compare it, and the economic second moment that the paper seeks to restrict.

Textual analysis in finance increasingly uses language models to encode
heterogeneous firm information in a common representation space.
Finance-specific models such as FinBERT improve sentiment analysis and
information extraction from financial text \citep{huang_finbert_2023}.
Building on these representations, more recent applications use
language-model outputs and embeddings to forecast returns and study how
markets process news \citep{chen_expected_2026,lopezlira_chatgpt_2026}.
These applications largely map individual documents into classifications,
scores, or expected-return predictions.
Text-based network industries instead measure firm similarity from 10-K
product descriptions \citep{hoberg_phillips_2016}; their embedding-based
successor supplies the product-market comparison used in our classification
tests \citep{hoberg2025scope}.

Textual similarity also carries information about financial dependence.
\citet{box2018qualitative} shows that similarity in firm-specific newswire
content predicts subsequent return correlation beyond historical
comovement and conventional firm characteristics, and evaluates its use
in minimum-variance portfolios.
\citet{dyer2024disclosure} document a corresponding relation for mandatory
financial disclosures, provide evidence consistent with disclosure
similarity influencing subsequent comovement, and connect pairwise
similarity to future market beta.
These findings motivate examining how heterogeneous firm information
relates to shared risk.
We retain each firm's collection of article representations as a distribution
and relate separation between these distributions to systematic covariance
under explicit transmission conditions.

Optimal transport supplies the geometry for that comparison.
Word Mover's Distance treats word embeddings as empirical distributions and
compares texts by the cost of transporting probability mass through semantic
space \citep{kusner_word_embeddings_2015}.
This construction shows how a collection of language-model representations
can become a probability measure whose geometry reflects distances in the
underlying representation space.
We extend this distributional comparison to the systematic covariance
supported by firms' factor exposures.

The economic foundation is Kantorovich's allocation problem: distribute a
given supply across destinations to meet their requirements at minimum cost
\citep{kantorovich_translocation_1958}.
Its dual expresses the same optimum through supporting shadow valuations,
connecting feasible allocations to the prices that support them
\citep{panaretos_invitation_2020}.
For proportional distance costs, the valuation restriction has a familiar
spatial-arbitrage interpretation: a price difference cannot exceed the cost
of moving the good between locations.
Quadratic costs instead connect to the rising marginal costs of adjustment
used in neoclassical investment theory \citep{hayashi_marginal_q_1982}.
These interpretations give the distributional comparison economic structure:
both the amount of mass reassigned and the cost of each reassignment matter.
Our application aligns distributions of reported firm characteristics;
in exposure space, quadratic cost also links that alignment exactly to
common-shock covariance through polarization.

The factor-model literature supplies the financial object to which that
geometry is applied: systematic covariance generated by exposure to common shocks.
In the Arbitrage Pricing Theory of \citet{ross_arbitrage_1976}, returns load
on pervasive factors and no-arbitrage restrictions discipline expected returns.
Focusing instead on second moments, \citet{chamberlain_rothschild_1983} study
approximate factor structure through the spectrum of the return covariance operator.
Characteristic-based factor-loading models treat exposures as functions of
firm descriptors
\citep{rosenberg1974extra,connor2007semiparametric,connor2012efficient}, and
the time-varying-beta literature allows those loadings to change
\citep{ghysels_stable_1998}.
Instrumented principal components use observed characteristics to model
dynamic loadings jointly with latent factors
\citep{kelly_characteristics_2019}.
Proxy-selection results further caution that proxy factors inherit
approximate pricing relations only under maintained factor structure and
additional rank conditions
\citep{reisman1992reference,shanken1992current,middleton2000deriving}.
Taken together, this literature distinguishes factor exposures, their
covariance implications, and the additional conditions required for
expected-return restrictions.
Our target is systematic covariance rather than expected returns because the
alignment of common-shock exposures determines how systematic risk aggregates
across firms without requiring a new no-arbitrage pricing claim.
Relative to these models, the paper changes the representation of firm
characteristics from a single descriptor vector to a probability law and asks
how distance between two such laws restricts the covariance their induced
exposures can support.

A complementary finance literature organizes risk through continuous spaces,
asset-indexed fields, and economically meaningful notions of proximity.
Random-field representations have been used for the term structure
\citep{kennedy_term_1994,goldstein_term_2000,santa_clara_sornette_2001}.
Related constructions model equity returns with a stochastic string
\citep{distaso_string_2024} and liquidity with an infinite-dimensional field
\citep{lototsky_liquidity}.
Continuum-of-assets models instead place the continuum over securities
\citep{werner1985,al1995decomposition,khan2001asymptotic}.
Spatial and network factor models encode economic proximity through
geographic, supplier--customer, or news co-mention links
\citep{menzly_market_2010,fernandez_spatial_2011,scherbina_economic_2013,bera_spatial_2016,kou_asset_2018,schwenkler_network_2020,ge_news-implied_2023,gao2025high}.
These approaches show that economic proximity can organize common risk.
Yet spatial and network models typically encode proximity with a discrete
relation matrix, while characteristic-based models represent each firm by a
finite descriptor or loading vector; neither formulation represents within-firm
heterogeneity as a distribution.

We instead represent each firm's text-derived characteristics as an empirical
probability measure and compare firms using quadratic Wasserstein distance.
Maintained restrictions on a common characteristic-to-exposure map transfer
bounds on that distance to latent exposure space.
There, polarization converts the minimum expected squared distance between
exposure laws into the attainable ceiling on pairwise systematic covariance.
Industry classifications summarize economic activity through discrete labels,
whereas estimated factor loadings summarize return sensitivity to a selected
set of shocks.
The distributional representation instead retains heterogeneous reported
information within each firm and permits a conditional covariance restriction
to be stated before selecting its realized dependence with another firm.
The empirical comparisons assess the distance association conditional on
sector and finer industry matching and on 10-K product-market similarity.
The next section formalizes this bridge and derives the resulting one-sided
envelope for the covariance ceiling.

\section{Economic transmission and the covariance envelope}\label{sec:methodology}

\subsection{Theoretical framework}

The framework converts distance between distribution-valued firm
characteristics into a conditional restriction on systematic covariance.
To make that argument transparent, the construction separates
characteristic-side information, the mechanism that transmits characteristics
into exposures, and the joint arrangement of the resulting exposure draws.
For firm $i$, the law $C_i$ records a distribution over risk-relevant
locations rather than collapsing heterogeneous information to one vector.
A common map sends draws from these laws to factor exposures, while a
Lipschitz restriction controls how much the map can distort characteristic distance.

Because the restriction concerns systematic covariance, the return
representation is written to accommodate either a finite or an infinite
collection of common risk directions.
A real separable Hilbert space generalizes the familiar Euclidean loading
space: its inner product measures alignment and its norm measures magnitude.
Let $\mathcal H$ be such a space, let $F_t\in\mathcal H$ be a centered factor
innovation with covariance operator $\Gamma$, let $e_t^i$ be a centered
idiosyncratic component, and write $\widetilde r_t^i$ for firm $i$'s centered return:
\begin{equation}\label{eq:hilbert-return}
  \widetilde r_t^i = \inner{B_i,F_t}_{\mathcal H}+e_t^i.
\end{equation}
The return equation contains two familiar sources of randomness: $F_t$
represents shocks common to firms, whereas $e_t^i$ represents the centered
firm-specific return component.
The random element $B_i$ adds a distinct source of variation by allowing a
firm's loading on common shocks to follow an exposure law.
To connect that loading law to distribution-valued characteristics, let
$B_i=T_i(X_i,U_i)$, where $X_i\sim C_i$ is a characteristic draw and $U_i$
allows the same characteristic to map to different exposures.
Thus $X_i$ represents heterogeneity within the firm's characteristic law,
while $U_i$ represents transmission variation conditional on a
characteristic; both differ from the return innovations $F_t$ and $e_t^i$.
Equivalently, a firm-specific stochastic kernel
$K_i(x,\cdot)=\mathcal L(T_i(x,U_i)\mid X_i=x)$ maps
characteristics to exposure laws and $P_i=C_i\,K_i=\mathcal L(B_i)$.
The common restriction below concerns the benchmark component of these maps;
the complete transmission maps may differ across firms.
The deterministic map and additive-slack cases are special cases of this construction.

Transport comparisons require the firm laws to share a characteristic domain
whose metric supplies the cost of moving between risk-relevant locations.
Let $(\Omega,d_\Omega)$ be a metric space of semantic risk locations with
Borel $\sigma$-algebra $\mathcal B$.
Fix any $\omega_0\in\Omega$ and define the admissible characteristic domain
\begin{equation}\label{eq:characteristic-domain}
  \mathcal P_1(\Omega)
  :=\left\{C:\ C\text{ is a Borel probability measure on }\Omega,\quad
  \int_\Omega d_\Omega(\omega,\omega_0)\,C(d\omega)<\infty\right\}.
\end{equation}
The definition is independent of the chosen base point by the triangle inequality.

The object to be bounded is the covariance of systematic components, so
exposures are first expressed in the factor-risk geometry induced by $\Gamma$.
Under a coupling $\pi$ of two exposure laws, define
\begin{equation}\label{eq:hilbert-systematic-covariance}
  \kappa_\pi=\E_\pi\inner{Z_i,Z_j}_{\mathcal H},
  \qquad Z_i:=\Gamma^{1/2}B_i.
\end{equation}
Thus $\Gamma$ weights factor directions by their contribution to systematic
risk, and $Z_i$ measures exposure in the covariance-relevant geometry.
Unlike $X_i$ or $U_i$, the coupling $\pi$ is not a further marginal shock; it
specifies how draws from two already-defined exposure laws occur jointly.

The envelope itself concerns systematic covariance, whereas its
interpretation using total returns requires a separate return bridge.
Under that bridge, the joint exposure draw is independent of the centered
factor innovation, whose covariance is $\Gamma$.
Residuals are square-integrable and centered, are orthogonal to both firms'
systematic returns, and have zero cross-firm covariance under the same joint law.
These conditions equate total-return covariance with $\kappa_\pi$.
Total-return variance additionally includes residual variance, which enters
the empirical return-distance calculation below.

\subsection{From economic characteristics to covariance}\label{sec:random-exposure}

The economic link starts with how firms earn revenue, incur costs, and
allocate risk through contracts.
Consider two airlines facing an unexpected increase in fuel prices.
Fuel intensity determines the immediate cost pressure, while competition and
ticket contracts constrain how much can be passed into fares.
Fuel hedges can offset part of the cash-flow loss; debt commitments affect
how the remaining loss falls on equity.
Managers choose routes, fleets, fares, and hedging arrangements subject to
contracts and adjustment costs.
Investors value the resulting expected cash flows and discount-rate risks.
A factor loading summarizes the sensitivity of returns to a specified
common shock, rather than the frequency with which that shock is mentioned.

Within a stable operating regime, comparable economic roles and contractual
arrangements motivate comparable responses to common shocks.
Differences in hedging, leverage, and operating arrangements motivate
firm-specific departures.
This reasoning motivates a shared benchmark relationship, called the common
carrier below, between economic characteristics and risk sensitivities.
The relevant characteristics must describe economic roles: an oil producer
and an airline can discuss the same commodity while responding in opposite
directions to its price.
Characteristic-based loading models likewise relate firm descriptors to
factor sensitivities through shared functions \citep{kelly_characteristics_2019}.

Characteristic distance can restrict covariance only if differences in
characteristics survive as differences in systematic exposures.
We therefore allow a shared response, stochastic variation, and bounded
firm-specific departures, and state explicitly how much similarity the
relationship must preserve.

Quadratic transport and covariance both require second moments, so this
subsection strengthens the characteristic domain accordingly.
Let $\Omega$ be a Polish metric space with its Borel $\sigma$-algebra, let
$\mathcal H$ be the exposure Hilbert space, and let $\Gamma:\mathcal
H\to\mathcal H$ be a bounded, positive self-adjoint operator.
Its unique bounded positive square root is denoted by $\Gamma^{1/2}$.
Write
\begin{equation}\label{eq:random-p2-domain}
  \mathcal P_2(\Omega)
  :=\left\{C\in\mathcal P_1(\Omega):
  \int_\Omega d_\Omega(\omega,\omega_0)^2\,C(d\omega)<\infty\right\}.
\end{equation}
The lower transfer bound will need to pull a coupling on the carrier image
back to characteristic space, which motivates the inverse condition stated next.
The common carrier $t:\Omega\to\mathcal H$ is Borel measurable, and
$t_\Gamma:=\Gamma^{1/2}\circ t$ is a Borel embedding with a Borel inverse on its image.
Each stochastic map $T_i:\Omega\times\mathcal U\to\mathcal H$ is Borel
measurable, and the maintained integrability conditions put the induced
exposure laws in $\mathcal P_2(\mathcal H)$.

The assumption below allows two controlled departures from exact common transmission.
Economically, $L$ measures how faithfully characteristic differences survive
as systematic-exposure differences: $L=1$ preserves distances, whereas $L>1$
permits multiplicative compression or expansion.
The bound $\tau_i$ permits an additive firm-level departure from the common
carrier in the same risk units.
This slack is a transmission tolerance rather than article-measurement error,
and the transmission draw $U_i$ that generates it remains distinct from the
cross-firm coupling $\pi$ used to join the firm-level exposure laws.
\begin{assumption}[Common carrier with bounded random departures]
  \label{ass:random-common-map}
  Let $C_i,C_j\in\mathcal P_2(\Omega)$ and let $t:\Omega\to\mathcal H$ be
  the common carrier. Assume that, for some $L\ge1$,
  \begin{equation}\label{eq:random-bilipschitz}
    L^{-1}d_\Omega(x,y)\le\norm{t_\Gamma(x)-t_\Gamma(y)}_{\mathcal H}
    \le Ld_\Omega(x,y).
  \end{equation}
  If $X_i\sim C_i$ and $U_i$ is a conditional transmission draw,
  define the exposure random element
  \begin{equation}\label{eq:random-exposure-pushforward}
    B_i=T_i(X_i,U_i)
    =t(X_i)+\varepsilon_i(X_i,U_i),
    \qquad \norm{\Gamma^{1/2}\varepsilon_i(x,u)}_{\mathcal H}\le\tau_i,
    \qquad P_i:=\mathcal L(B_i).
  \end{equation}
  The slack maps $\varepsilon_i$ are Borel measurable, and the preceding
  integrability condition ensures that $P_i$ has a finite
  $\Gamma$-weighted second moment. The maps, the common Lipschitz scale $L$,
  and the deviation bounds $\tau_i$ are maintained over the analyzed regime.
  Both scales are stated in risk coordinates:
  equation~\ref{eq:random-bilipschitz}
  constrains $\Gamma^{1/2}t$ rather than $t$ alone, and $\tau_i$ bounds the
  deviation after the same factor weighting. All transport quantities below
  use this factor-risk geometry.
\end{assumption}

The two sides of the restriction have different economic content.
The upper bound imposes bounded sensitivity: within the maintained regime,
small characteristic changes cannot produce arbitrarily large changes in
risk-weighted exposure.
For example, with contracts and financing held within a stable regime,
a modest change in fuel intensity or cost pass-through can change the
fuel-price sensitivity gradually.
Smooth cash-flow and valuation responses with bounded sensitivities motivate
this local approximation; distress, contractual thresholds, and regime
changes can undermine it.
The lower bound separately requires that distinctions represented by the
characteristic metric survive as exposure differences.
It can fail when characteristics distinguish economically irrelevant features,
when hedging offsets differences in operating exposure, or when different
activities have offsetting systematic exposures.
Thus two airlines with different fuel requirements could nevertheless have
similar net fuel-price exposure after hedging.
The lower restriction therefore selects characteristics whose differences
remain economically relevant after firms' risk-management decisions.

A finite-dimensional example makes the sufficient conditions explicit.
Take $\Omega$ to be a compact subset of $\mathbb R^k$ with Euclidean distance,
and let the carrier be $t(x)=b_0+Ax$, where $b_0\in\mathcal H$ and
$A:\mathbb R^k\to\mathcal H$ is linear.
This affine benchmark represents a first-order approximation to the response
of factor loadings to economic characteristics within a fixed operating regime;
firm-specific departures and approximation error enter $\varepsilon_i$.
If, on the span of admissible characteristic differences, constants
$0<m\le M<\infty$ satisfy
\[
  m\norm{h}\le\norm{\Gamma^{1/2}Ah}_{\mathcal H}\le M\norm{h},
\]
then equation~\ref{eq:random-bilipschitz} holds with
$L=\max\{1,M,m^{-1}\}$.
The upper constant limits amplification; the positive lower constant excludes
characteristic directions that have no systematic-risk effect.
Bounding the risk-weighted approximation and firm-specific departures by
$\tau_i$ supplies the remaining transmission condition.

The economic magnitude of $L$ depends on the fixed characteristic units and
the factor-risk geometry.
Common amplification or compression enters $L$, while differences from that
carrier enter $\tau_i$.
For example, $L=1.25$ permits carrier distances between $80\%$ and $125\%$
of characteristic distances before additive slack is allowed.
The empirical grid varies $L$ and $\tau_i$ as scenario inputs to show how
much transmission uncertainty the restriction tolerates.
The relative scale of characteristic and exposure units is part of each scenario.
Estimating that conversion would require economic exposure information in
addition to normalized articles and returns.
The constant $L$ bounds distortion over all admissible characteristic pairs;
the regression below summarizes an average association.

Once the transmission relationship fixes the exposure laws, comparing their
possible joint arrangements is a Kantorovich transport problem.
A coupling specifies which exposure draws occur together while preserving
each firm's marginal law.
The transport cost measures how far those draws are apart, and minimizing it
finds their closest feasible alignment.
This gives the cost specification an economic role before it enters the
covariance calculation.

With cost proportional to displacement, W$_1$ measures the least average
distance required for alignment.
Its dual asks for the largest difference in average valuation across the
two distributions, subject to a unit bound on valuation changes per unit
distance \citep{panaretos_invitation_2020}.
In a goods market with symmetric transport costs, that bound rules out
profitable spatial arbitrage: buying at one location, transporting, and
reselling at another cannot yield a positive margin.
Here the dual valuations give an economic reading of distributional
separation in the chosen characteristic units.
The characteristic-to-exposure relationship remains the separate
transmission assumption above.

Quadratic cost gives larger adjustments increasing marginal weight.
This is familiar from the Tobin--Hayashi investment mechanism: firms balance
the marginal value of installed capital against its purchase and marginal
installation costs, so rising adjustment costs encourage gradual investment
\citep{hayashi_marginal_q_1982}.
A smooth adjustment penalty with positive curvature at its no-adjustment
minimum has a local quadratic form; curvature sets the relative cost of
adjustment in different directions.
For a mass-preserving reconfiguration over a unit horizon, W$_2^2$ is the
least integrated squared adjustment velocity
\citep{benamou_brenier_2000,santambrogio_optimal_transport_2015}.
W$_2$ is its square root, expressed in displacement units.
\Cref{sec:transport-economic-interpretation} records the dual and dynamic
formulations and their normalization.

The quadratic specification has an additional role specific to the financial
question here: the same squared distance enters the inner-product identity
that determines common-shock covariance.
We state this identity before deriving the covariance envelope.

\begin{lemma}[Polarization Identity]\label{lem:polarization}
  For any $x,y$ in a real inner-product space,
  \begin{equation}\label{eq:polarization}
    \inner{x,y}
    =\frac12\left(\norm{x}^2+\norm{y}^2-\norm{x-y}^2\right).
  \end{equation}
\end{lemma}

Polarization supplies the specific link between quadratic transport and
covariance: its squared ground cost is exactly the quantity that
equation~\ref{eq:polarization} converts into an inner product.
Minimizing squared transport cost over couplings therefore maximizes
covariance over couplings, which gives W$_2$ a theoretical role not shared by
the alternative distances used later as empirical comparisons.

Three quantities summarize the pair in factor-risk coordinates: the marginal
second moment $v_i$, the coupling-specific covariance $\kappa_\pi$, and the
minimum squared exposure distance $W_{2,\Gamma}^2$.
For $Z_i=\Gamma^{1/2}B_i$, write
\begin{equation}\label{eq:random-covariance-objects}
  v_i:=\E\norm{Z_i}^{2},\qquad
  \kappa_\pi:=\E_\pi\inner{Z_i,Z_j},\qquad
  W_{2,\Gamma}^{2}(P_i,P_j):=\inf_\pi
  \E_\pi\norm{Z_i-Z_j}^{2}.
\end{equation}
The infimum ranges over couplings of $P_i$ and $P_j$ after the covariance
geometry induced by $\Gamma$.
Because the marginal second moments remain fixed as the coupling changes,
define the candidate upper endpoint
\begin{equation}\label{eq:random-covariance-envelope}
  \bar\kappa_{ij}:=\frac12\left(v_i+v_j-W_{2,\Gamma}^{2}(P_i,P_j)\right).
\end{equation}

The central result establishes that this endpoint is the attainable
covariance ceiling and identifies the gap between that ceiling and any
particular coupling.

\begin{theorem}[Random-exposure covariance envelope]
  \label{thm:random-exposure-envelope}
  Under \Cref{ass:random-common-map}, for every coupling $\pi$ of the exposure
  laws,
  \begin{equation}\label{eq:random-exposure-gap}
    \kappa_\pi
    =\bar\kappa_{ij}-\frac12\Delta_\pi,
    \qquad
    \Delta_\pi:=\E_\pi\norm{Z_i-Z_j}^{2}
    -W_{2,\Gamma}^{2}(P_i,P_j).
  \end{equation}
  The excess is nonnegative,
  \begin{equation}\label{eq:random-exposure-excess-nonneg}
    \Delta_\pi\ge0 ,
  \end{equation}
  and consequently $\bar\kappa_{ij}$ is the greatest covariance attainable by
  a coupling of the two exposure laws, with every realized coupling satisfying
  $\kappa_\pi\le\bar\kappa_{ij}$.
\end{theorem}

\begin{proof}
  Polarization in $\mathcal H$ gives
  $\norm{Z_i-Z_j}^2=\norm{Z_i}^2+\norm{Z_j}^2-2\inner{Z_i,Z_j}$ pointwise.
  Integrating against $\pi$ and using that its marginals are $P_i$ and $P_j$
  replaces the first two terms by $v_i$ and $v_j$, which depend on the
  marginals alone, so $\E_\pi\norm{Z_i-Z_j}^2=v_i+v_j-2\kappa_\pi$.
  Subtracting $W_{2,\Gamma}^2(P_i,P_j)$ and dividing by two gives
  equation~\ref{eq:random-exposure-gap}. That step is an identity in
  $W_{2,\Gamma}^2$ and holds for any reference cost placed there;
  equation~\ref{eq:random-exposure-excess-nonneg} is the separate consequence
  of that particular reference cost being the infimum of the same expectation
  over the same coupling set. The ceiling is attained rather than merely
  approached because
  the quadratic cost is lower semicontinuous and the coupling set is weakly
  compact, so an optimal plan exists and realizes $\bar\kappa_{ij}$
  \citep[Ch.~1]{panaretos_invitation_2020}.
\end{proof}

Holding $v_i$ and $v_j$ fixed, a larger latent exposure distance lowers the
greatest systematic covariance that the two laws can attain.
The transport excess $\Delta_\pi$ measures how much the realized joint
arrangement exceeds the minimum transport cost and, equivalently, how far its
covariance falls below the ceiling.

To characterize the entire feasible covariance set rather than only its
ceiling, the same argument is applied to the reflected law $\check P_j:=(-I)_\#P_j$.

\begin{theorem}[Reflected lower endpoint]
  \label{thm:random-exposure-lower-endpoint}
  Under \Cref{ass:random-common-map}, the least systematic covariance
  attainable over the coupling set is
  \begin{equation}\label{eq:random-exposure-floor}
    \underline\kappa_{ij}
    =\frac12\left(W_{2,\Gamma}^{2}(P_i,\check P_j)-v_i-v_j\right).
  \end{equation}
\end{theorem}

\begin{proof}
  Reflection $(z_i,z_j)\mapsto(z_i,-z_j)$ bijects couplings of
  $(P_i,P_j)$ with couplings of $(P_i,\check P_j)$, preserves second moments,
  and negates covariance.
  Applying \Cref{thm:random-exposure-envelope} to the reflected pair and
  negating its upper endpoint gives equation~\ref{eq:random-exposure-floor}.
\end{proof}

The reflected endpoint describes the most oppositely aligned feasible
arrangement of the two exposure laws.
Mixtures of the endpoint-attaining couplings preserve the marginals and
attain every covariance between $\underline\kappa_{ij}$ and $\bar\kappa_{ij}$.
The main application uses the upper endpoint: greater separation limits the
scope for aligned systematic exposure.

The covariance envelope still depends on the latent distance
$W_{2,\Gamma}(P_i,P_j)$, which is not observed in the application.
The next step is therefore to use \Cref{ass:random-common-map} to transfer
bounds from characteristic-side distance, later proxied by article
embeddings, to exposure-side distance.
The following metric bracket compares each exposure law with the
pushforward of its characteristic law through the common carrier.
A complementary squared-cost bound for finite clouds is given in
\Cref{app:finite-support-transfer}.

\begin{corollary}[Metric transfer bracket]\label{cor:random-w2-metric}
  Under \Cref{ass:random-common-map}, with
  $G_{ij}:=W_{2,d_\Omega}(C_i,C_j)$, $\tau_{ij}:=\tau_i+\tau_j$, and
  $(x)_+:=\max\{x,0\}$,
  \begin{align}
    \left(L^{-1}G_{ij}-\tau_{ij}\right)_+
    &\le W_{2,\Gamma}(P_i,P_j)
    \label{eq:random-w2-lower-metric}\\
    W_{2,\Gamma}(P_i,P_j)
    &\le LG_{ij}+\tau_{ij}.
    \label{eq:random-w2-upper-metric}
  \end{align}
\end{corollary}

\begin{proof}
  Write $W_{2,\Gamma}(\mu,\nu):=W_2\bigl((\Gamma^{1/2})_\#\mu,
  (\Gamma^{1/2})_\#\nu\bigr)$, so that every leg below is measured after the
  same factor weighting and $(\Gamma^{1/2})_\#(t_\#C_i)=(t_\Gamma)_\#C_i$.
  Coupling $B_i$ with $t(X_i)$ through the shared draw $X_i$ gives
  $W_{2,\Gamma}^2(P_i,t_\#C_i)
  \le\E\norm{\Gamma^{1/2}\varepsilon_i(X_i,U_i)}^2\le\tau_i^2$, and pushing an
  optimal plan for $(C_i,C_j)$ forward through $t_\Gamma$ gives
  $W_{2,\Gamma}(t_\#C_i,t_\#C_j)\le LG_{ij}$ by the upper
  half of equation~\ref{eq:random-bilipschitz}. The $W_2$ triangle inequality
  on $\mathcal P_2(\mathcal H)$ chains the three legs into
  equation~\ref{eq:random-w2-upper-metric}.
  For the lower endpoint, let $\gamma$ be an arbitrary coupling of
  $(t_\Gamma)_\#C_i$ and $(t_\Gamma)_\#C_j$. Its marginals are supported on
  $t_\Gamma(\Omega)$, so $\gamma$ is supported on
  $t_\Gamma(\Omega)^2$. The Borel inverse on this image therefore defines
  \[
    \widetilde\gamma
    :=\bigl((t_\Gamma)^{-1},(t_\Gamma)^{-1}\bigr)_\#\gamma.
  \]
  Its marginals are $C_i$ and $C_j$. The lower half of
  equation~\ref{eq:random-bilipschitz} gives, after integration,
  \[
    G_{ij}^{2}
    \le L^{2}\int
    \norm{t_\Gamma(x)-t_\Gamma(y)}^{2}\,d\widetilde\gamma(x,y).
  \]
  Taking infima over $\gamma$ yields
  $W_{2,\Gamma}(t_\#C_i,t_\#C_j)\ge L^{-1}G_{ij}$. The same
  triangle inequality subtracts the two slack legs, and the positive part is
  taken because a Wasserstein distance is nonnegative.
\end{proof}

The bracket makes the role of the transmission assumptions transparent.
When $L=1$ and $\tau_i=\tau_j=0$, characteristic distance is preserved in
factor-risk coordinates; as multiplicative distortion or additive slack
increases, the lower endpoint falls and the upper endpoint rises.

To pass this exposure-distance bracket into the covariance ceiling, write
\begin{equation}\label{eq:random-w2-bracket}
  w^-_{ij}:=\left(L^{-1}W_{2,d_\Omega}(C_i,C_j)-\tau_i-\tau_j\right)_+,
  \qquad
  w^+_{ij}:=LW_{2,d_\Omega}(C_i,C_j)+\tau_i+\tau_j,
\end{equation}
and substitute \Cref{cor:random-w2-metric} into
equation~\ref{eq:random-covariance-envelope} to obtain the conditional
interval implied by characteristic-side distance and the maintained scales:
\begin{equation}\label{eq:random-kappa-bracket}
  \frac12\left(v_i+v_j-(w^+_{ij})^2\right)
  \le \bar\kappa_{ij}
  \le \frac12\left(v_i+v_j-(w^-_{ij})^2\right).
\end{equation}
The interval $[w^-_{ij},w^+_{ij}]$ is the range of latent exposure distances
allowed by the maintained transmission inputs.
Its lower endpoint has the direct economic implication used below: a larger
$w^-_{ij}$ tightens the covariance ceiling because characteristics that
cannot be brought close under any admissible transmission leave less room for
shared systematic exposure.
The empirical analysis uses article distance to represent characteristic
separation and evaluates this restriction across $(L,\tau)$ scenarios.

\section{Data and empirical design}\label{sec:empirical-design}

The application represents firms by distributions of financial articles and
relates their separation to return co-movement.
We first construct the sample and its article geometry, then estimate the
distance association with firm effects and classification controls.
A sensitivity grid applies the covariance restriction across transmission scenarios.
\Cref{sec:data-provenance} documents retrieval, filtering, and data provenance.

\subsection{Sample and return construction}\label{sec:sample-and-window}

Firm-level distribution comparisons require sufficient text and common
coverage across the fixed window to construct a common-size empirical measure
for every firm, which motivates a coverage-conditioned sample rather than a
broad but uneven news universe.
The news universe is a prespecified, Nasdaq-100-based frame of
\ppvalue{n-firms-news} Nasdaq-listed firms.
Each retained firm had at least \ppvalue{annual-raw-article-floor} raw articles
in every calendar year from
\ppvalue{sample-year-start} through \ppvalue{sample-year-end} in Nasdaq's
public per-symbol news archive.
Within the fixed frame, articles satisfying the date-window and
minimum-body-length filters are ordered by URL hash, and up to
\ppvalue{n-samples-cap} are retained over the pooled window.
This coverage rule gives each firm a sufficiently supported empirical
distribution on a common observation window.
\Cref{tab:sample-descriptives-sectors} reports the sector composition,
per-firm article counts, return volatilities, and within- versus cross-sector
W$_2$ distances.
\Cref{tab:sample-flow} records the resulting flow from source records to
estimation dyads; the full retrieval and provenance protocol is in
\Cref{sec:data-provenance}.


\begin{table}[H]

\centering

\small
\begin{tabularx}{\linewidth}{Xrrrrr}
\toprule
Sector & $n$ & Articles & Volatility & $W_2$ within & $W_2$ cross \\
\midrule
Communication Services & 11 & 128.0 & 0.0239 & 1.042 & 1.060 \\
Consumer Cyclical & 18 & 128.0 & 0.0281 & 1.033 & 1.053 \\
Consumer Defensive & 7 & 128.0 & 0.0176 & 1.010 & 1.058 \\
Energy & 1 & 128.0 & 0.0386 & \textemdash & 1.045 \\
Financial Services & 1 & 128.0 & 0.0272 & \textemdash & 1.070 \\
Healthcare & 14 & 128.0 & 0.0233 & 1.025 & 1.065 \\
Industrials & 7 & 128.0 & 0.0240 & 1.030 & 1.057 \\
Technology & 38 & 128.0 & 0.0265 & 0.998 & 1.050 \\
Utilities & 3 & 128.0 & 0.0159 & 0.882 & 1.058 \\
\midrule
All sectors & 100 & 128.0 & 0.0251 & 1.008 & 1.056 \\
\bottomrule
\end{tabularx}

\caption{Per-sector sample composition: firm count, mean article count per firm, mean daily return volatility, and mean within- versus cross-sector pairwise rooted $W_2$ distance. Articles are subsampled to a fixed per-firm cap, so counts are near-constant by construction. A dash indicates a statistic that is undefined for that sector (e.g.\ within-sector distance for a singleton sector).}

\label{tab:sample-descriptives-sectors}

\end{table}

Two features of that composition motivate later design choices.
First, mean daily return volatility differs markedly across sectors, whereas
the primary return chord distance is constructed from correlations and is
therefore invariant to firm-specific return scale.
Symmetric firm effects instead absorb stable endpoint heterogeneity.
Second, mean within-sector W$_2$ distance is smaller than the corresponding
cross-sector distance in every sector holding more than one firm, so
article-embedding distance encodes sector information before any return is used.
The sector diagnostic in \Cref{sec:geometric-representation} quantifies that
pattern, and both the pooled ladder and the classification-adjusted
firm-effects regressions probe it.

To measure co-movement independently of firm-specific return scale, the
outcome is constructed from correlations of daily simple returns.
Daily simple returns are Yahoo Finance adjusted-close ratios minus one.
Return-dependent tests use the exact complete-case matrix of
\ppvalue{return-panel-n-observations} dates shared by
\ppvalue{return-panel-n-tickers} priced firms, spanning
\ppvalue{return-panel-date-start} through \ppvalue{return-panel-date-end}.
Nullable observations are removed before matrix construction, with no
interpolation or zero filling.
The primary analysis relates news and returns over the common
\ppvalue{sample-year-start}--\ppvalue{sample-year-end} window.
Return alignment and sector-metadata details are documented in
\Cref{sec:data-provenance}.

\subsection{Article-embedding representation and distance}
\label{sec:distance-construction}\label{sec:article-proxy-model}

The empirical bridge requires a common representation of heterogeneous
articles while retaining each firm as a distribution rather than collapsing
its text to a single point.
Embeddings provide that measurement technology by mapping documents into a
shared numerical space in which semantic proximity can be compared.
An embedding model is a neural network that maps a text document to a
fixed-length numerical vector---its \emph{embedding}.
Documents discussing similar topics are placed nearby in the resulting
high-dimensional space, whereas documents on unrelated subjects are placed
farther apart.
A firm's collection of vectors can therefore be treated as an empirical
distribution in semantic space.

A bi-encoder processes each document independently into a reusable vector,
giving all firm distributions a common geometry for Wasserstein comparison.

Let $C_i^\ast$ denote firm $i$'s latent economic characteristic law,
corresponding to $C_i$ in the theory, and let $Q_i$ be the law of its reported
content after encoding by $\phi$.
With $N_i$ observed articles, the empirical distribution is
\begin{equation}\label{eq:empirical-embedding-measure}
  \widehat Q_i
  :=\frac{1}{N_i}\sum_{d=1}^{N_i}\delta_{\phi(d)}.
\end{equation}
Distances between $\widehat Q_i$ supply the empirical measure of firm separation.
Their economic interpretation depends on how reporting and encoding preserve
information about $C_i^\ast$.

Financial news is an economically selected information product.
Investors seek information that helps them assess expected cash flows,
discount rates, and shared risks.
Information producers choose which firms and developments to investigate
under reporting costs, demand for coverage, and reputational constraints.
A story about fuel costs may therefore be useful to investors in several
airlines, while the value of investigating one firm's hedge contracts is
more concentrated.
When producing information is costly but distributing it to additional users
is cheap, widely demanded information can be supplied at a lower cost per
user.
Veldkamp's information-market model shows how these incentives encourage
shared information acquisition and can themselves generate asset-price
comovement \citep{veldkamp_information_markets_2006}.
This provides a reason for economic content to appear repeatedly across
firms' news, while also identifying a source of comovement beyond common
operating exposures.

Reports can supply overlapping evidence about a common economic state.
News of demand for one premium smartphone, for example, informs an assessment
of category demand as well as the producer's competitive position.
The firms' economic roles and investors' prior expectations determine how
that information translates into valuation.

News producers also select stories for audience appeal and professional
reasons.
Survey evidence indicates that financial journalists value accuracy,
timeliness, and depth, but also select provocative companies and personalities
for coverage \citep{call_meet_press_2022}.
Reputation incentives can coexist with persistent reporting bias
\citep{gentzkow_media_bias_2006}.
Article frequencies therefore reflect both economic activity and the
incentives to report it.

A stylized selection mechanism clarifies the measurement requirement.
Before text generation and encoding, let $s_i(x)\ge0$ be the measurable
reporting intensity of economic characteristic $x$, with
$0<\int s_i\,dC_i^\ast<\infty$.
The distribution of characteristics selected for reporting is then
\begin{equation}\label{eq:news-selection}
  \widetilde C_i^\ast(dx)
  :=\frac{s_i(x)C_i^\ast(dx)}{\int s_i(u)C_i^\ast(du)}.
\end{equation}
If $s_i(x)=a_i s(x)$ for a positive firm-specific multiplier $a_i$,
normalization removes that multiplier but retains the common topic weighting
$s(x)$.
Consequently, article frequencies represent reported economic salience rather
than revenue shares or economic event probabilities.
Comparable reporting incentives motivate cross-firm measurement, but even a
common topic weighting need not preserve economic distances.
Through text generation and encoding, the coverage law $Q_i$ is generated
from the selected law $\widetilde C_i^\ast$ rather than from $C_i^\ast$ itself.
The empirical interpretation therefore requires the following premise about
the distortions introduced by selection and representation, separately from
the transmission bounds.

\begin{assumption}[Measurement Premise: Article-Embedding
  Proxy]\label{ass:article-proxy}
  After symmetric firm margins and the declared controls, distortions from
  media selection, article generation, duplicate handling, text extraction,
  and embedding do not covary with return distance strongly enough to create
  or reverse the sign of the latent association.
\end{assumption}

This premise links reported content to economic characteristics.
The firm-specific kernel $K_i$ then maps those characteristics to exposure
law $P_i$ under the transmission conditions in \Cref{ass:random-common-map}.
The tolerance $\tau_i$ measures departure from the common economic response;
measurement distortion is addressed separately by \Cref{ass:article-proxy}.

The distributions pool persistent attributes and transient events over the
sample window, providing a common-window summary of reported firm information.
The 10-K comparison in \Cref{sec:classification-controls} assesses how this
news geometry relates to firms' descriptions of their own businesses.

We use Qwen3-Embedding-8B for the headline article representation
\citep{qwen3_embedding_2025}.
Under pinned serving parameters, each full article body $d$ maps to a vector
$\bm{e}_d=\phi(d)\in\R^{\ppvalue{embed-dim}}$, an atom of
$\widehat Q_i$ in equation~\ref{eq:empirical-embedding-measure}.
Holding article counts fixed keeps the number and weight of atoms comparable
across firms rather than combining semantic comparisons with unequal
empirical support sizes.
Within each firm, articles are ordered lexicographically by their
stable URL hash.
All full-width encoder comparisons use the first $m=128$ articles under this
ordering; for the headline encoder we additionally use $m\in\{32,64,128\}$.
Let $x_{ia}$ denote the row-normalized embedding of selected article $a$, and
let $S_m$ be the set of permutations of $\{1,\ldots,m\}$.
The primary statistic is
\begin{equation}\label{eq:empirical-wasserstein-2}
  \widehat W_{2,m}(i,j)
  = \left(\min_{\pi\in S_m}\frac{1}{m}\sum_{a=1}^{m}
  \norm{x_{ia}-x_{j,\pi(a)}}_2^{2}\right)^{1/2}.
\end{equation}
The minimization pairs every article from one firm with one article from the
other and finds the smallest root-mean-square semantic displacement, so W$_2$
is small when two firms' article topics can be aligned closely even if no
individual article is shared.
The fixed-W$_2$ diagnostic uses the common $m=128$ unit-chord Qwen3
representation.
The complete W$_1$ and energy definitions are in
Appendix~\ref{sec:representation-diagnostics} as alternative distance summaries.

\subsection{Distance association and inference}
\label{sec:testable-restrictions}
\label{sec:primary-estimator}\label{sec:empirical-testing}
\label{sec:inference-design}

We regress return distance on article-embedding distance after absorbing
stable firm-level heterogeneity.
Its outcome is the return chord distance
\begin{equation}\label{eq:chord-distance}
  \hat{D}_{ij} = \sqrt{2(1 - \hat{\rho}_{ij})},
\end{equation}
where $\hat{\rho}_{ij}$ is the sample Pearson correlation of daily returns
(equation~\ref{eq:sample-correlation} in \Cref{sec:data-provenance}).
On this scale, $D_{ij}=0$ denotes perfectly correlated returns,
$D_{ij}=\sqrt{2}$ denotes zero correlation, and $D_{ij}=2$ denotes perfectly
negative correlation.
Larger chord distance therefore means weaker return co-movement.

The regression is defined over the observed undirected pairs $i<j$:
\begin{equation}\label{eq:dyadic-fixed-effects}
  D_{ij}=\mu+a_i+a_j+\beta_X X_{ij}+z_{ij}^{\top}\gamma+u_{ij},
  \qquad 1\le i<j\le N.
\end{equation}
Here $N=\ppvalue{n-firms}$, $\mu$ is the intercept, $X_{ij}$ is the
article-embedding regressor with coefficient $\beta_X$, $z_{ij}$ is the
vector of secondary controls with coefficient vector $\gamma$, and $u_{ij}$
is the dyadic error.
The terms $a_i+a_j$ are symmetric additive firm effects.
The restriction $i<j$ counts each unordered pair once.
The normalization omits one firm-effect indicator; changing the omitted
ticker does not change $\widehat\beta_X$.
The canonical specification sets $X_{ij}=W^{(2)}_{ij}$ and $z_{ij}=0$,
yielding the fixed-W$_2$ slope $\beta_{W_2}$.
To test incremental association beyond classifications in this same
firm-effects design, we set $z_{ij}=S_{ij}$, where $S_{ij}$ indicates that
the two firms share a sector.
A finer adjustment sets $z_{ij}=(S_{ij},I_{ij})^{\top}$, where $I_{ij}$
indicates a shared industry, and a within-industry regression retains only
pairs with $I_{ij}=1$ and absorbs symmetric firm effects.
In each case, partial $R^2$ measures the reduction in residual sum of
squares from adding W$_2$ to a restricted model retaining all other controls.
For example, the sector-adjusted baseline includes both firm effects and
$S_{ij}$.

\begin{hypothesis}[Reduced-form distance association]
  \label{hyp:monotonicity}
  For the theory-aligned arm, the reported two-sided regression test is
  \begin{equation}\label{eq:dyadic-association-hypothesis}
    H_0:\beta_{W_2}=0
    \qquad\text{against}\qquad
    H_1:\beta_{W_2}\ne0.
  \end{equation}
  We separately hypothesize $\beta_{W_2}>0$: greater article distance is
  associated with greater return distance after symmetric firm effects.
  The covariance envelope motivates this question; the sign of the average
  association is an empirical hypothesis.
\end{hypothesis}

Dyads sharing a firm may be dependent, so the cross-sectional information is
governed by the number of firms rather than the number of pairs.
Firms are therefore the resampling units, and symmetric additive firm effects
absorb stable endpoint heterogeneity.
Each multinomial node-bootstrap draw assigns a count to every firm and
weights dyad $(i,j)$ by the product of its endpoint counts to account for
dependence among pairs sharing a firm.
Percentile intervals and the finite-draw-corrected two-sided sign tail use
the same 1,999-draw schedule for every reported specification.

The residual representation, partial-$R^2$ definition, nuisance-rank details,
six-rung estimator audit, exact weights, rank checks, functional-form
diagnostics, and influence results are reported in
Appendix~\ref{sec:dyadic-diagnostics}.

A later-window specification holds the
\ppvalue{sample-year-start}--\ppvalue{sample-year-end} article-embedding
W$_2$ matrix fixed and rebuilds return chord distances and symmetric firm
effects for the 2023--2026 complete-case panel.
It uses the same dyads, estimator, bootstrap schedule, and two-sided null,
measuring persistence of the association across return windows.
The features are constructed retrospectively; encoder and classification
timing are documented in \Cref{sec:data-provenance}.

\subsection{Covariance sensitivity}\label{sec:covariance-sensitivity}

The covariance envelope provides a second use of article distance:
assessing candidate covariances under specified transmission conditions.
We apply the return bridge in \Cref{sec:methodology}, expressing exposures
and residuals in standardized total-return units.
Then $\rho_{ij}=\kappa_\pi$, but a firm's unit return variance is
$v_i+\sigma_{e_i}^2=1$, where $\sigma_{e_i}^2$ is its residual variance.
Let $q_{ij}:=\sigma_{e_i}^2+\sigma_{e_j}^2\ge0$.
Writing $\Delta_{ij}$ for the pair's realized transport excess, subtracting
optimal exposure cost from observed return distance gives
\begin{equation}\label{eq:total-return-excess-correction}
  H_{ij}:=c^{\mathrm{actual}}_{ij}-W_{2,\Gamma}^2(P_i,P_j)
  =\Delta_{ij}+q_{ij},
  \qquad c^{\mathrm{actual}}_{ij}:=2-2\rho_{ij}.
\end{equation}
Residual variance contributes additively to this diagnostic.
The transmission bracket gives
\begin{equation}\label{eq:random-transport-excess-bracket}
  c^{\mathrm{actual}}_{ij}-(w^+_{ij})^2
  \le H_{ij}
  \le c^{\mathrm{actual}}_{ij}-(w^-_{ij})^2.
\end{equation}
We classify an interval as \emph{nonnegative} when its lower endpoint is
nonnegative, as a \emph{violation} when its upper endpoint is negative, and
as \emph{unresolved} when it spans zero.
Since $H_{ij}=\Delta_{ij}+q_{ij}\ge0$ under the joint measurement,
transmission, and return conditions, a negative upper endpoint flags
inconsistency in the plug-in scenario for the pair.
Recovering an interval for systematic excess $\Delta_{ij}$ additionally
requires subtracting residual variance $q_{ij}$ from both endpoints.

For this numerical scenario, we identify characteristic distance $G_{ij}$
with the Qwen3 $m=128$ article distance at the declared normalization and
impose the transmission restrictions on exposures in standardized-return units.
This numerical identification is additional to the association-level
measurement premise in \Cref{ass:article-proxy}.
The calculation uses estimated return moments
$c^{\mathrm{actual}}_{ij}=2-2\hat\rho_{ij}$.
The nine-cell grid varies multiplicative distortion
$L\in\{1,1.25,1.5\}$ and firm-level slack
$\tau_i=\tau_j\in\{0,0.05,0.10\}$ around the isometric, zero-slack benchmark.
These scenario inputs determine how much transmission uncertainty the
covariance restriction admits.

\subsection{Representation sensitivity}\label{sec:representation-ladder}

Representation checks ask which measurement choices sustain the distance
association, holding the firm sample, return matrix, estimator, and bootstrap
schedule fixed.
Supplementary lexical comparisons use firm-level TF--IDF cosine distance
and W$_2$ between distributions of TF--IDF article vectors.
Four additional W$_2$ comparisons use base encoders pretrained with causal
or masked language modeling; these supplementary comparisons use only
2018--2022 returns (\Cref{tab:text-controls}).
Qwen capacity variants compare model size, including a comparison at a common
output width; BGE changes the encoder architecture and training pipeline.
Nested leading-coordinate representations, trained to remain usable at
smaller widths through Matryoshka representation learning, assess how
compression changes the measured relationship.
W$_1$ substitutes proportional for quadratic transport cost, while energy
provides a different summary of the same article distributions.
The published ChronoBERT checkpoints for 2017--2024 assess sensitivity to the
encoder's reported training-information cutoff within a common architecture
\citep{he_chronobert_2025}.
Later checkpoints continue training earlier ones, so these are cumulative
annual-training comparisons, not matched-compute causal tests of leakage.
\Cref{sec:representation-diagnostics} gives construction details, paired
contrasts and their benchmark-calibrated equivalence rule, and
article-sampling checks.

\section{Empirical results and discussion}\label{sec:results}

Article distance is positively associated with return distance across both
return windows and after classification and product-market controls.
We present that evidence first, examine its stability across representations,
and then assess the informativeness of the covariance restriction.

\subsection{Article distance and return co-movement}
\label{sec:post-window-dyadic}

Greater article-embedding distance is associated with weaker return
co-movement after symmetric firm effects.
Across representative dyads in the interquartile range, a pair one standard
deviation farther apart in news content has a fitted return correlation
roughly eight points lower.
Return-chord distance maps deterministically to the familiar correlation
scale through \eqref{eq:chord-distance}, because $\hat\rho=1-\hat D^2/2$.
At the median dyad, a chord distance of \ppnum[3]{chord-median} corresponds
to a return correlation of \ppnum[3]{chord-rho-median}.
A one-standard-deviation increase in W$_2$ moves the implied correlation to
\ppnum[3]{chord-rho-after-median}, a change of
\ppnum[3]{chord-delta-rho-median}.
The corresponding changes are \ppnum[3]{chord-delta-rho-q25} at the
lower-quartile dyad and \ppnum[3]{chord-delta-rho-q75} at the
upper-quartile dyad.

\Cref{tab:w2-primary-results} reports the same symmetric-firm-effect
estimator for both return windows.
In the shared 2018--2022 window, the focal coefficient is
\ppnum[3]{confound-w2-coef} return-chord units per unit of W$_2$, with
standard error \ppnum[3]{confound-w2-se} and 95\% interval
[\ppnum[3]{confound-w2-ci-low}, \ppnum[3]{confound-w2-ci-high}].
A one-standard-deviation increase in W$_2$ corresponds to
\ppnum[3]{confound-w2-effect-one-sd} return-chord units, and W$_2$ accounts
for \ppnum[3]{confound-partial-r2} of residual variation relative to
symmetric firm effects.
The positive estimate is consistent with the directional expectation in
\Cref{hyp:monotonicity}.

The positive association persists in the later return window.
The exercise fixes the 2018--2022 article-embedding geometry before
recomputing return chord distances and firm effects for
\ppvalue{oos-return-panel-date-start} through
\ppvalue{oos-return-panel-date-end}.
The coefficient rises to \ppnum[3]{oos-dyadic-w2-coef}, with 95\% interval
[\ppnum[3]{oos-dyadic-w2-ci-low}, \ppnum[3]{oos-dyadic-w2-ci-high}], and the
standardized effect is \ppnum[3]{oos-dyadic-w2-effect-one-sd}.
Thus the same article geometry is associated with co-movement in both
return windows.

\begin{table}[H]
\centering
\small
\begin{tabularx}{\linewidth}{Xrrrrrrr}
\toprule
Return window & Firms & Dyads & $\hat\beta_{W_2}$ & 95\% CI & $p$ & $\Delta D$/1 SD & Partial $R^2$ \\
\midrule
2018--2022 returns & 100 & 4950 & 1.874 & [1.516, 2.250] & 0.001 & 0.069 & 0.332 \\
2023--2026 returns & 100 & 4950 & 1.945 & [1.516, 2.454] & 0.001 & 0.071 & 0.314 \\
\bottomrule
\end{tabularx}
\caption{Primary symmetric-firm-effect association. The text-side rooted $W_2$ matrix is fixed across rows. The first row relates the 2018--2022 return chord distance to that geometry; the second uses the later 2023--2026 return panel. Intervals and two-sided sign-tail probabilities use the multinomial node bootstrap. $\Delta D$/1 SD standardizes by the full-sample dyadic standard deviation of $W_2$; partial $R^2$ is relative to symmetric additive firm effects.}
\label{tab:w2-primary-results}
\end{table}

The residual-space diagnostics show how the fitted relation is distributed
across pairs and how it responds to changes in functional form and firm composition.
By the Frisch--Waugh--Lovell result, Panel A of
\Cref{fig:dyadic-postfit-diagnostics} residualizes both return distance and
W$_2$ with respect to the symmetric firm effects, so the slope through those
residuals equals the canonical coefficient.
Panels B and C assess functional form and single-firm influence.

\begin{figure}[H]
  \centering
  \import{images/}{dyadic_postfit_diagnostics\ppfiguresuffix.pgf}
  \caption{Post-fit diagnostics for the canonical
    symmetric-firm-effects specification.
    Panel A plots firm-effect-residualized return chord distance against
    residualized W$_2$; the fitted line uses all dyads, while equal-count bin
    averages are descriptive.
    Panel B plots raw W$_2$ against the residualized outcome and overlays the
    prespecified quadratic sensitivity.
    Panel C reports the ten largest leave-one-firm-out coefficient shifts.
    The residual relation is positive, and the coefficient remains
    positive after removing each firm in turn.
  }
  \label{fig:dyadic-postfit-diagnostics}
\end{figure}

The positive association is stable across these checks.
Classification controls next assess how much of it remains after accounting
for familiar measures of pair-level economic similarity.

\subsection{Sector, industry, and product-market
comparisons}\label{sec:classification-controls}

\Cref{tab:classification-controls} adds sector matching directly to the
primary symmetric-firm-effects design.
The coefficient remains positive under this combined adjustment.
The W$_2$ coefficient is \ppnum[3]{classification-sector-coefficient}, with
bootstrap standard error \ppnum[3]{classification-sector-se} and 95\%
interval [\ppnum[3]{classification-sector-ci-low},
\ppnum[3]{classification-sector-ci-high}].
For a one-standard-deviation increase in distance, the fitted chord contrast
is \ppnum[3]{classification-sector-effect-one-sd}, corresponding to a change
of \ppnum[3]{classification-sector-delta-correlation} in implied correlation
at the median pair.
Distance explains \ppnum[3]{classification-sector-partial-r2} of the
variation remaining after both firm effects and sector matching.

Adding same-industry matching, or restricting the fit to within-industry
pairs, leaves the coefficient positive with a 95\% interval excluding zero;
\Cref{sec:empirical-robustness} discusses these finer fits.

Industry labels are coarse relative to product-market similarity measured
from firms' own disclosures.
The last two rows of \Cref{tab:classification-controls} therefore restrict the
panel to the \ppnum[0]{etnic-n-firms} firms covered by the Hoberg--Phillips
ETNIC network and add its 10-K business-description similarity, averaged over
fiscal years \ppvalue{etnic-year-start}--\ppvalue{etnic-year-end}
\citep{hoberg2025scope}.
News W$_2$ and 10-K business-description distance have correlation
\ppnum[3]{etnic-matrix-correlation-w2}
(permutation $p\le\ppnum[4]{etnic-matrix-correlation-pvalue-max}$ under
  \ppnum[0]{etnic-matrix-correlation-permutations} joint row-and-column
firm relabelings).
After removing sector matching from both matrices, their descriptive
correlation is \ppnum[3]{etnic-matrix-correlation-w2-partial-sector}.
This agreement provides convergent evidence across news and disclosure text.
Both sources reflect reporting choices, and journalists also draw on company
disclosures.
Adding ETNIC similarity to firm effects and sector matching lowers the W$_2$
coefficient from \ppnum[3]{classification-etnic-roster-coefficient} to
\ppnum[3]{classification-etnic-coefficient}, with 95\% interval
[\ppnum[3]{classification-etnic-ci-low},
\ppnum[3]{classification-etnic-ci-high}].
Distance explains \ppnum[3]{classification-etnic-partial-r2} of the variation
remaining after firm effects, sector matching, and ETNIC similarity; ETNIC
similarity explains \ppnum[3]{classification-etnic-control-partial-r2} after
the other terms.
Both rows use the same firms and resampling draws, so the change reflects the
control rather than the sample.
The remaining association is consistent with additional economic content in
news, imperfect measurement of product similarity, and shared reporting
patterns.
The regression establishes its incremental contribution conditional on the
recorded ETNIC measure.

\begin{table}[H]
\centering\small
\setlength{\tabcolsep}{3pt}
\begin{tabularx}{\linewidth}{>{\raggedright\arraybackslash}Xrrrrrrr}
\toprule
Controls/sample & Firms & Dyads & $\hat\beta_{W_2}$ & 95\% CI & $\Delta D$/1 SD & $\Delta\rho$ & Partial $R^2$ \\
\midrule
FE + same sector & 100 & 4950 & 1.721 & [1.338, 2.129] & 0.063 & -0.072 & 0.227 \\
FE + same sector + same industry & 100 & 4950 & 1.555 & [1.167, 1.967] & 0.057 & -0.065 & 0.184 \\
Within industry, FE & 86 & 182 & 1.444 & [0.530, 2.442] & 0.054 & -0.050 & 0.309 \\
\midrule
ETNIC firms: FE + same sector & 81 & 3240 & 1.765 & [1.297, 2.280] & 0.063 & -0.070 & 0.228 \\
\quad + ETNIC similarity & 81 & 3240 & 1.314 & [0.858, 1.880] & 0.047 & -0.052 & 0.112 \\
\bottomrule
\end{tabularx}
\caption{Classification controls in the primary symmetric-firm-effects design, 2018--2022 returns. FE denotes symmetric additive firm effects. The first two rows add pair-level matching indicators to the full panel; the third retains only pairs sharing an industry. Labels come from the undated sample-roster snapshot, so these are retrospective controls. Intervals are 95\% multinomial node-bootstrap percentile intervals. Full-panel rows reuse the main schedule; the within-industry row resamples its incident firms, retaining only within-industry edges. $\Delta D$/1 SD uses each row's distance standard deviation; $\Delta\rho$ translates that effect at the row's median chord distance using $\rho=1-D^2/2$ and is a fitted contrast. Partial $R^2$ is $1-\mathrm{SSE}_{\mathrm{full}}/\mathrm{SSE}_{\mathrm{restricted}}$, where the restricted model removes only $W_2$, retaining all controls. The within-industry row changes the sample and has a distinct baseline. The last two rows restrict the panel to firms covered by the Hoberg--Phillips ETNIC network and share one resampling schedule; ETNIC is the mean 10-K business-description similarity over the fiscal years of the return window.}
\label{tab:classification-controls}
\end{table}

\subsection{Stability across representations}
\label{sec:representation-sensitivity}

Consistent with \citet{box2018qualitative}, lexical similarity is associated
with return comovement.
This association also appears when firms are represented as distributions
of TF--IDF article vectors: greater W$_2$ distance is associated with greater
return distance after symmetric firm effects, with a 95\% node-bootstrap
interval excluding zero and a partial $R^2$ of
\ppnum[3]{text-controls-tfidf-document-partial-r2}
(\Cref{tab:text-controls}).
These results support the empirical usefulness of distribution-valued firm
representations and Wasserstein comparison beyond the headline
language-model representation.

The positive distance association persists across encoder families,
embedding widths, alternative distance summaries, and reported
training-information vintages.
Moderate compression preserves the association, whereas the more severe
compression steps attenuate it, showing where the loss of semantic
coordinates weakens the measured relation.
The Qwen capacity variants and the BGE representation retain positive
associations, and their paired differences from the relevant references are
not distinguishable after adjustment.
W$_1$ nearly preserves the canonical dyad ordering, whereas energy retains a
positive association under a materially different ordering.
The close W$_1$/W$_2$ agreement may partly reflect the geometry imposed by
unit normalization.
This maps embeddings to the unit sphere, retaining direction while removing
vector length, as in many contrastive representation systems
\citep{wang_isola_hypersphere_2020}.
If the distances relevant to article matching occupy a narrow range,
proportional and quadratic costs can produce similar orderings.
This suggests a measurement question: whether representations designed to
preserve economically meaningful variation in both direction and magnitude
would distinguish the two costs more sharply.
\Cref{sec:representation-diagnostics} develops this interpretation as a
hypothesis about the representation.

The ChronoBERT annual comparisons meet the inherited benchmark-calibrated
equivalence rule for every pair, but some differences are distinguishable
from zero after vintage-family Holm adjustment
(\Cref{tab:w2-vintage-contrasts}).
Later training cutoffs neither monotonically strengthen nor monotonically
weaken the standardized association.
Thus the paired differences are bounded under this robustness margin,
not exactly zero; cumulative training also prevents a causal interpretation
as the effect of later information alone.
\Cref{tab:w2-representation-sensitivity} reports the annual cells, with
construction details and article-sampling checks in
\Cref{sec:representation-diagnostics}.

\subsection{Covariance sensitivity}\label{sec:fixed-w2-test}

The association establishes the empirical relevance of article distance.
The sensitivity grid addresses the next question: how tightly must economic
transmission be specified for that distance to restrict covariance informatively?
\Cref{tab:random-exposure-w2} applies the interval in
\Cref{eq:random-transport-excess-bracket} across nine transmission scenarios.
The zero-slack isometric cell $(L,\tau)=(1,0)$ gives nonnegative intervals for
\ppnum[3]{confound-w2-l1-tau0-coverage} of observed dyads and violations for
\ppnum[3]{confound-w2-l1-tau0-violation}, resolving every interval's sign.
With $(L,\tau)=(1,0.05)$, the unresolved share rises to
\ppnum[3]{confound-w2-l1-tau0p05-unresolved}; at $(1,0.10)$ it reaches
\ppnum[3]{confound-w2-l1-tau0p1-unresolved}.
Even modest firm-specific departures therefore widen the interval enough to
leave most comparisons unresolved.

\begin{table}[H]
\centering
\small
\begin{tabularx}{\linewidth}{rr>{\raggedleft\arraybackslash}X>{\raggedleft\arraybackslash}X>{\raggedleft\arraybackslash}X}
\toprule
$L$ & $\tau_i=\tau_j$ & Nonnegative & Unresolved & Violation \\
\midrule
1.00 & 0.00 & 0.707 [0.625, 0.782] & 0.000 [0.000, 0.000] & 0.293 [0.218, 0.375] \\
1.00 & 0.05 & 0.349 [0.262, 0.438] & 0.560 [0.496, 0.627] & 0.091 [0.055, 0.135] \\
1.00 & 0.10 & 0.076 [0.042, 0.115] & 0.898 [0.863, 0.932] & 0.026 [0.012, 0.046] \\
1.25 & 0.00 & 0.015 [0.006, 0.028] & 0.960 [0.941, 0.976] & 0.025 [0.012, 0.045] \\
1.25 & 0.05 & 0.000 [0.000, 0.001] & 0.994 [0.987, 0.998] & 0.006 [0.001, 0.013] \\
1.25 & 0.10 & 0.000 [0.000, 0.000] & 0.999 [0.996, 1.000] & 0.001 [0.000, 0.004] \\
1.50 & 0.00 & 0.000 [0.000, 0.000] & 0.996 [0.991, 0.999] & 0.004 [0.001, 0.009] \\
1.50 & 0.05 & 0.000 [0.000, 0.000] & 0.999 [0.997, 1.000] & 0.001 [0.000, 0.003] \\
1.50 & 0.10 & 0.000 [0.000, 0.000] & 1.000 [1.000, 1.000] & 0.000 [0.000, 0.000] \\
\bottomrule
\end{tabularx}
\caption{Fixed-W2 total-return compatibility diagnostic. Intervals bracket $H_{ij}=\Delta_{ij}+q_{ij}$, where $q_{ij}$ is the residual-variance sum in standardized returns and $c^{\mathrm{actual}}_{ij}=2-2\rho_{ij}$. Nonnegative denotes a nonnegative lower endpoint; violation denotes a negative upper endpoint; unresolved denotes an interval spanning zero. A violation flags plug-in inconsistency with the scenario. Entries are shares of the 4,950 dyads; brackets are percentile intervals from the common multinomial node bootstrap. Article distance uses 128-article measures and unit-chord ground cost. $L$ and $\tau$ specify transmission distortion and firm-level slack.}
\label{tab:random-exposure-w2}
\end{table}

For an investment manager assessing a prospective holding, the workflow
begins by comparing its distribution of reported activities with those of
existing holdings, including firms outside its sector.
The analyst then reviews the underlying articles for shared customers,
suppliers, and contracts, and examines how pricing power, hedging, and
financing transmit those risks to equity returns.
Those economic judgments specify the transmission scenario and exposure
magnitudes used to assess candidate covariances against the ceiling.
The sensitivity grid shows how precise these inputs must be for that
assessment to discriminate among candidate covariances.
The neighbour paths in \Cref{sec:exploratory-geometry-outcomes} illustrate
the range of return behaviour that the economic review must explain.

\section{Conclusion}\label{sec:conclusion}

Representing firms by distributions preserves the heterogeneity of their
risk-relevant information and permits covariance restrictions to be stated
in terms of distances between probability laws.
For given exposure laws, quadratic Wasserstein geometry determines sharp
covariance endpoints.
A common transmission benchmark with bounded firm-specific departures
connects characteristic distance to the covariance ceiling.
Holding exposure magnitudes fixed, economic characteristics that remain
separated under admissible transmission leave less scope for shared
systematic risk.

The Nasdaq application gives this framework an observable counterpart.
A one-standard-deviation increase in article-embedding distance corresponds
to roughly eight fewer correlation points after symmetric firm effects.
The association persists across sector and industry controls, within
industries, after conditioning on 10-K product-market similarity, and in a
later return window.
These comparisons show that news distributions contain information about
co-movement beyond the recorded classifications and product-market measure.

The sensitivity grid identifies where the theoretical restriction is useful.
Tight transmission scenarios discriminate among covariance comparisons;
modest increases in distortion or firm-specific slack leave many comparisons
unresolved.
The practical value therefore lies in combining information about a firm's
reported activities with an explicit account of its exposure to common shocks.
Article distance directs attention to pairs worth examining, while the
envelope makes the role of contracts, hedging, financing, and transmission
uncertainty explicit in assessing their potential systematic covariance.

\section*{Funding}

The authors received no financial support for the research, authorship, or
publication of this article.

\section*{Disclosure statement}

The authors report no potential conflict of interest.

\bibliographystyle{plainnat}
\bibliography{refs}

\appendix
\section{Data construction and coverage}\label{sec:data-provenance}

The application uses a declared Nasdaq-100-based frame rather than a
reconstructed historical-membership panel. A firm enters the news frame only
if the raw Nasdaq per-symbol archive contains at least
\ppvalue{annual-raw-article-floor} articles in every
calendar year from \ppvalue{sample-year-start} through
\ppvalue{sample-year-end}. The screen ensures annual support for each empirical
article distribution but conditions the sample on media coverage.

The comparison in \Cref{tab:news-dataset-comparison} highlights the trade-off:
the public source permits researcher-controlled full-text measurement but
offers narrower market coverage than large proprietary archives.

\begin{table}[H]
  \centering
  \footnotesize
  \caption{Comparison of finance-news corpora used in text-implied network
    studies. ``Partial'' denotes a qualified or product-dependent property;
    vendor entity tags are not manual gold-standard annotations. The final row
  describes the source used in this study.}
  \label{tab:news-dataset-comparison}
  \begin{tabularx}{\linewidth}{@{}p{0.18\linewidth}
      >{\raggedright\arraybackslash}X
      >{\raggedright\arraybackslash}X
      >{\raggedright\arraybackslash}X
      >{\raggedright\arraybackslash}X
      >{\raggedright\arraybackslash}X
    >{\raggedright\arraybackslash}X@{}}
    \toprule
    Dataset and representative citation &
    English / major-market coverage &
    Full article body &
    Entity annotation &
    Scale / coverage &
    Access / reproducibility &
    Researcher text control \\
    \midrule
    RavenPack Equity---Dow Jones Edition
    \citet{ge_news-implied_2023} &
    Yes, English Dow Jones sources; the focal study uses U.S. large-cap
    equities &
    No, standard academic analytics files do not provide article bodies &
    Partial, high-quality vendor entity IDs but no manual gold labels &
    Yes, large commercial archive; 2004--2015 in the focal study &
    No, proprietary; WRDS entitlement required &
    No, vendor ontology and analytics limit raw-text experiments \\
    Thomson Reuters News Analytics (TRNA)
    \citet{scherbina_economic_2013} &
    Yes, English Reuters; broad international archive with a U.S. focal
    application &
    Partial, the analytics feed lacks the body; pair with Reuters archive or
    MRN &
    Partial, vendor firm tags but no manual gold labels &
    Yes, approximately 5.5 million stories from 1996--2012 &
    No, commercial access; analytics and archive entitlements may differ &
    Partial, rich metadata; body-level NLP requires a companion feed \\
    Raw Reuters financial-news corpus (Ding et al. corpus)
    \citet{schwenkler_network_2020} &
    Yes, English Reuters; U.S.-oriented firm sample &
    Yes, complete article text &
    Partial, automated NER by the authors and no manual gold-standard labels &
    Partial, 106{,}521 articles from 2006--2013 &
    No, the historical files were removed from public download for copyright
    reasons &
    Yes, researchers control NER, sentence rules, and embeddings \\
    RESET Financial Text Intelligent Analysis Platform + Juyuan
    \citet{ge_network_guided_covariance_2026} &
    No, Chinese-language financial news; focused on China A-shares &
    Partial, sentence-level use is documented; bulk export is licence-dependent &
    Partial, company mentions are available; the exact annotation pipeline is
    unclear &
    Yes, 1{,}138{,}247 relevant articles from 2006--2021; pre-2012 coverage is
    thinner &
    No, proprietary databases with uncertain bulk-export rights &
    Partial, passage and sentence definitions can vary; raw text may be
    restricted \\
    Nasdaq-hosted public per-symbol news archive (this study) &
    Yes, English syndicated news for Nasdaq-listed U.S. firms; not global &
    Yes, article bodies are retrieved and filtered by body length &
    Partial, ticker-level source assignment but no article-level vendor or
    manual entity labels &
    Partial, coverage-conditioned Nasdaq-100-based frame and fixed five-year
    window rather than a representative market sample &
    Yes, public pages are viewable and retrievable; provider terms and rate
    limits still apply &
    Yes, bodies can be independently extracted and embedded with a pinned
    open-weights model \\
    \bottomrule
  \end{tabularx}
\end{table}

\begin{table}[H]
  \centering
  \caption{Data construction and estimation flow. Counts are pipeline-generated;
    article selection uses stable URL-hash order and return dates use an exact
  complete-case rule. }
  \label{tab:data-provenance}
  \label{tab:sample-flow}
  \begin{tabularx}{\linewidth}{lXr}
    \toprule
    Stage & Source and rule & Retained \\
    \midrule
    Raw article records & Public Nasdaq per-symbol news archive &
    \ppnum[0]{corpus-raw-records} \\
    Deduplicated universe & Identifier-level deduplication,
    \ppvalue{corpus-year-min}--\ppvalue{corpus-year-max} &
    \ppnum[0]{corpus-articles-total} \\
    Analysis window & Articles dated
    \ppvalue{sample-year-start}--\ppvalue{sample-year-end} &
    \ppnum[0]{corpus-articles-window} \\
    Source pool & Up to \ppvalue{n-samples-cap} articles per news-frame firm &
    \ppnum[0]{source-pool-total-articles} \\
    News firms & Prespecified coverage-qualified frame &
    \ppvalue{n-firms-news} \\
    Priced firms & Intersection with local adjusted-close histories &
    \ppvalue{n-firms} \\
    Estimation articles & \ppnum[0]{sample-desc-articles-per-firm} per priced firm &
    \ppnum[0]{sample-desc-total-articles} \\
    Return dates & Common dates from \ppvalue{return-panel-date-start} to
    \ppvalue{return-panel-date-end} &
    \ppvalue{return-panel-n-observations} \\
    Estimation dyads & Unordered priced pairs $i<j$ & \ppvalue{n-pairs} \\
    \bottomrule
  \end{tabularx}
\end{table}

Identifier-level deduplication removes
\ppnum[0]{corpus-duplicates-dropped} repeated records. After the fixed window
and deterministic cap, full article bodies are embedded independently with the
pinned encoder and row-normalized before exact balanced W$_2$ construction.
Daily simple returns are adjusted-close ratios minus one; nullable observations
are removed before matrix construction, with no interpolation or zero filling.
The news frame contains \ppnum[0]{n-firms-news} firms, and its intersection with
compatible local return histories retains \ppnum[0]{n-firms} priced firms.

The canonical priced W$_2$ roster used by the return-dependent plots and
estimates is listed in \Cref{tab:sample-universe}.


\begin{longtable}{lll}
\caption{Canonical empirical roster: ticker symbol, firm name, and sector label from Nasdaq summary metadata for the firms in the canonical priced $W_2$ geometry, sorted by sector then symbol.}
\label{tab:sample-universe} \\
\toprule
Symbol & Name & Sector \\
\midrule
\endfirsthead
\multicolumn{3}{c}{\tablename\ \thetable\ (continued)} \\
\toprule
Symbol & Name & Sector \\
\midrule
\endhead
\midrule
\endfoot
\bottomrule
\endlastfoot
BIDU & Baidu, Inc. & Communication Services \\
CHTR & Charter Communications, Inc. & Communication Services \\
CMCSA & Comcast Corporation & Communication Services \\
EA & Electronic Arts Inc. & Communication Services \\
GOOG & Alphabet Inc. & Communication Services \\
MTCH & Match Group Inc. & Communication Services \\
NFLX & Netflix Inc. & Communication Services \\
NTES & NetEase, Inc. & Communication Services \\
SIRI & Sirius XM Holdings Inc. & Communication Services \\
TMUS & T-Mobile US Inc. & Communication Services \\
TTWO & Take-Two Interactive Software, Inc. & Communication Services \\
BKNG & Booking Holdings Inc. & Consumer Cyclical \\
EBAY & eBay Inc. & Consumer Cyclical \\
EXPE & Expedia Group Inc. & Consumer Cyclical \\
HAS & Hasbro Inc. & Consumer Cyclical \\
JD & JD.com Inc. & Consumer Cyclical \\
LULU & Lululemon Athletica Inc. & Consumer Cyclical \\
MAR & Marriott International Inc. & Consumer Cyclical \\
MAT & Mattel Inc. & Consumer Cyclical \\
MELI & MercadoLibre Inc. & Consumer Cyclical \\
NCLH & Norwegian Cruise Line Holdings Ltd. & Consumer Cyclical \\
ORLY & O'Reilly Automotive Inc. & Consumer Cyclical \\
ROST & Ross Stores, Inc. & Consumer Cyclical \\
SBUX & Starbucks Corporation & Consumer Cyclical \\
TCOM & Trip.com Group Limited & Consumer Cyclical \\
TRIP & Tripadvisor, Inc. & Consumer Cyclical \\
TSCO & Tractor Supply Company & Consumer Cyclical \\
ULTA & Ulta Beauty Inc. & Consumer Cyclical \\
WYNN & Wynn Resorts Limited & Consumer Cyclical \\
COST & Costco Wholesale Corporation & Consumer Defensive \\
DLTR & Dollar Tree Inc. & Consumer Defensive \\
KDP & Keurig Dr Pepper Inc. & Consumer Defensive \\
KHC & The Kraft Heinz Company & Consumer Defensive \\
MDLZ & Mondelez International, Inc. & Consumer Defensive \\
MNST & Monster Beverage Corporation & Consumer Defensive \\
PEP & PepsiCo Inc. & Consumer Defensive \\
FANG & Diamondback Energy, Inc. & Energy \\
PYPL & PayPal Holdings Inc. & Financial Services \\
ALGN & Align Technology Inc. & Healthcare \\
AMGN & Amgen Inc. & Healthcare \\
AZN & AstraZeneca PLC & Healthcare \\
BIIB & Biogen Inc. & Healthcare \\
BMRN & BioMarin Pharmaceutical Inc. & Healthcare \\
DXCM & DexCom, Inc. & Healthcare \\
GILD & Gilead Sciences Inc. & Healthcare \\
HOLX & Hologic, Inc. & Healthcare \\
IDXX & IDEXX Laboratories, Inc. & Healthcare \\
ILMN & Illumina, Inc. & Healthcare \\
INCY & Incyte Corporation & Healthcare \\
ISRG & Intuitive Surgical Inc. & Healthcare \\
REGN & Regeneron Pharmaceuticals Inc. & Healthcare \\
VRTX & Vertex Pharmaceuticals Incorporated & Healthcare \\
AAL & American Airlines Group Inc. & Industrials \\
CSX & CSX Corporation & Industrials \\
FAST & Fastenal Company & Industrials \\
HON & Honeywell International Inc. & Industrials \\
JBHT & J.B. Hunt Transport Services, Inc. & Industrials \\
PCAR & PACCAR Inc & Industrials \\
UAL & United Airlines Holdings Inc. & Industrials \\
ADBE & Adobe Inc. & Technology \\
ADI & Analog Devices, Inc. & Technology \\
ADP & Automatic Data Processing Inc. & Technology \\
ADSK & Autodesk, Inc. & Technology \\
AKAM & Akamai Technologies, Inc. & Technology \\
AMAT & Applied Materials Inc. & Technology \\
AMD & Advanced Micro Devices Inc. & Technology \\
ASML & ASML Holding N.V. & Technology \\
AVGO & Broadcom Inc. & Technology \\
CDNS & Cadence Design Systems, Inc. & Technology \\
CHKP & Check Point Software Technologies Ltd. & Technology \\
CSCO & Cisco Systems Inc. & Technology \\
CTSH & Cognizant Technology Solutions Corporation & Technology \\
ENPH & Enphase Energy Inc. & Technology \\
FISV & Fiserv, Inc. & Technology \\
FTNT & Fortinet Inc. & Technology \\
INTC & Intel Corporation & Technology \\
INTU & Intuit Inc. & Technology \\
KLAC & KLA Corporation & Technology \\
LRCX & Lam Research Corporation & Technology \\
MCHP & Microchip Technology Incorporated & Technology \\
MRVL & Marvell Technology Inc. & Technology \\
MU & Micron Technology Inc. & Technology \\
NTAP & NetApp, Inc. & Technology \\
NVDA & NVIDIA Corporation & Technology \\
NXPI & NXP Semiconductors N.V. & Technology \\
OKTA & Okta, Inc. & Technology \\
PANW & Palo Alto Networks Inc. & Technology \\
PAYX & Paychex, Inc. & Technology \\
QCOM & QUALCOMM Incorporated & Technology \\
SNPS & Synopsys, Inc. & Technology \\
STX & Seagate Technology Holdings plc & Technology \\
SWKS & Skyworks Solutions, Inc. & Technology \\
TEAM & Atlassian Corporation & Technology \\
TXN & Texas Instruments Incorporated & Technology \\
WDAY & Workday Inc. & Technology \\
WDC & Western Digital Corporation & Technology \\
ZS & Zscaler Inc. & Technology \\
AEP & American Electric Power Company, Inc. & Utilities \\
EXC & Exelon Corporation & Utilities \\
XEL & Xcel Energy Inc. & Utilities \\
\end{longtable}

\Cref{tab:sample-descriptives-sectors}, reported in the body, gives coverage,
mean daily return volatility, and within- versus cross-sector W$_2$ distances.
The estimation design selects the same number of articles for each priced
firm. Energy and Financial Services
are singleton sectors in the current roster, so their within-sector quantities
are reported as undefined rather than zero. The imbalance and the missing
timestamp on Nasdaq sector metadata limit sector-based interpretation.
The same roster supplies \ppnum[0]{classification-n-industries} finer
industry labels for the classification checks in
\Cref{tab:classification-controls}.
Every priced firm has both labels, but the roster records no effective dates
or classification history.
The labels serve as retrospective classification controls.

The product-market comparison uses the Hoberg--Phillips ETNIC-All pair
scores, a preliminary release keyed by Compustat identifier, for fiscal years
\ppvalue{etnic-year-start}--\ppvalue{etnic-year-end}
\citep{hoberg_phillips_2016,hoberg2025scope}.
The preliminary files were downloaded on 23 September 2026; the frozen
sample extract and matching crosswalk are preserved with checksums in the
replication materials.
A curated ticker--CIK--Compustat crosswalk links the roster to the network.
Foreign issuers filing annual reports on Form 20-F are excluded by the
network's construction, firms without a crosswalk link are omitted, and firms
without co-observed pairs are dropped, leaving \ppnum[0]{etnic-n-firms} firms
whose every pair is scored.
ETNIC trains a separate doc2vec model on each year's business descriptions
and scores each pair by the cosine similarity of its boilerplate-purged
representations.
Pair similarity here is the mean of those scores over co-observed fiscal
years.
The fiscal-year average overlaps the return window and serves as a
retrospective product-market control; filing dates are not used for alignment.

The headline encoder is \texttt{qwen/qwen3-embedding-8b}, a fixed
semantic-retrieval model with no documented
training cutoff.
The later-window comparison uses articles dated before the evaluation
returns and embeddings computed retrospectively with that model.

The sample correlation underlying the chord distance in
equation~\ref{eq:chord-distance} is
\begin{equation}\label{eq:sample-correlation}
  \hat{\rho}_{ij}
  =
  \frac{
    \sum_{t=1}^{N_T} (r_t^i - \bar{r}^i)(r_t^j - \bar{r}^j)
  }{
    \sqrt{
      \sum_{t=1}^{N_T} (r_t^i - \bar{r}^i)^2
      \sum_{t=1}^{N_T} (r_t^j - \bar{r}^j)^2
    }
  },
\end{equation}
where $N_T$ is the number of common return dates and $\bar{r}^i$ is the
sample average of asset $i$'s returns, yielding a
$\ppvalue{n-firms} \times \ppvalue{n-firms}$ symmetric chord distance matrix
$D$ with zero diagonal.

\section{Empirical robustness and diagnostics}
\label{sec:empirical-robustness}

This appendix first audits the dyadic estimator and inference, then turns to
representation and finite article sampling, and finally considers the sector
projection and exploratory return-path display used in \Cref{sec:results}.

\subsection{Dyadic specification and inference audit}
\label{sec:dyadic-diagnostics}
\label{sec:resampling-formulas}

For the reported coefficient, let $X$ and $D$ stack $X_{ij}$ and $D_{ij}$ over
the observed dyads $i<j$, and let $F$ collect the intercept and one indicator
per non-omitted firm from the rank-cleaned point design. This nuisance design
is distinct from the secondary-control vector $z_{ij}$ in
equation~\ref{eq:dyadic-fixed-effects}, which is zero in the canonical
specification. Write $M_F=I-F(F^{\top}F)^{-}F^{\top}$, with
$\widetilde X=M_F X$ and $\widetilde D=M_F D$. The
Frisch--Waugh--Lovell identity is
\begin{equation}\label{eq:fwl-identity}
  \widehat\beta_X
  =\frac{\widetilde X^{\top}\widetilde D}{\widetilde X^{\top}\widetilde X}.
\end{equation}
In words, both axes are first stripped of their average firm-endpoint
components. The remaining slope therefore measures whether unusually distant
article distributions for a given pair coincide with unusually weak return
co-movement for that pair.

The nuisance block has \ppvalue{confound-fwl-nuisance-columns} columns and
omits the indicator for \texttt{\ppvalue{confound-fwl-omitted-firm}}.

The conditional, or partial, $R^2$ of the article-embedding regressor is
\begin{equation}\label{eq:partial-r2}
  R^2_{\mathrm{partial}}
  =1-\frac{\mathrm{SSE}_{\mathrm{FE}+X}}{\mathrm{SSE}_{\mathrm{FE}}},
\end{equation}
where $\mathrm{SSE}_{\mathrm{FE}}$ is the residual sum of squares from the
intercept and symmetric firm effects, and $\mathrm{SSE}_{\mathrm{FE}+X}$ is
from that model with $X$ added. It is defined only for the restricted
fixed-effect model and is therefore omitted on pooled rungs.
For the classification-adjusted estimates in
\Cref{tab:classification-controls}, replace $F$ by the joint nuisance block
containing firm effects and the included matching indicators or ETNIC
similarity.
The corresponding numerator and denominator in \eqref{eq:partial-r2} retain
those controls, so the partial $R^2$ isolates the contribution of $X$.
The within-industry fit uses the same construction on its restricted graph;
redundant endpoint indicators in disconnected industry groups are removed
by rank selection.

The finer classification fits in \Cref{tab:classification-controls} support
the sector-adjusted estimate.
Adding the same-industry indicator gives a W$_2$ coefficient of
\ppnum[3]{classification-industry-coefficient}, with 95\% interval
[\ppnum[3]{classification-industry-ci-low},
\ppnum[3]{classification-industry-ci-high}], and a partial $R^2$ of
\ppnum[3]{classification-industry-partial-r2} relative to the combined
firm-effects, sector, and industry baseline.
Restricting attention to within-industry pairs yields a coefficient of
\ppnum[3]{classification-within-industry-coefficient}, with 95\% interval
[\ppnum[3]{classification-within-industry-ci-low},
\ppnum[3]{classification-within-industry-ci-high}].
This interval is appreciably wider because the estimate uses only
\ppnum[0]{classification-within-industry-n-dyads} pairs among
\ppnum[0]{classification-within-industry-n-firms} incident firms.
Industries with only two or three firms supply no residual pair variation
after symmetric firm effects; the identifying sample comprises
\ppvalue{classification-identifying-industries} larger industries containing
\ppvalue{classification-identifying-firms} firms.

The six-rung W$_2$ ladder asks whether the association between
article-embedding distance and return distance is confined to raw sample
composition or remains after increasingly stringent adjustment for observable
and stable firm-level heterogeneity.
The sequence is increasingly demanding in the heterogeneity it absorbs,
although the final firm-effect specification changes the adjustment design
rather than simply adding another pooled control.
Specification 1 records the unadjusted W$_2$ association.
Specification 2 adds a same-sector indicator to address broad sector sorting;
because sector remains a coarse description of firm composition,
specification 3 adds the absolute difference in firms' mean log dollar
trading volume and an indicator that both return histories cover the
COVID-19 crisis window.
Specification 4 additionally declares the absolute difference and minimum
of the two firms' log article counts.
All firms cover the crisis window and have the same article count in this
sample, so the crisis indicator and both count controls are collinear with
the intercept or zero and are removed by rank selection.
Specifications 3 and 4 therefore have the same effective regressors
and estimates.
Article counts do not capture within-firm content dispersion, so
specification 5 adds the sum and absolute difference of that dispersion.
This decomposition diagnostic separates between-firm transport distance from
heterogeneity in the breadth of the two article clouds.
Because the pooled controls address only their measured dimensions,
specification 6 replaces them with symmetric firm effects that absorb stable
additive endpoint heterogeneity.
This sixth specification supplies the firm-effects benchmark; the preceding
five document the path from raw association to that design.
The classification tests in the main text extend the same firm-effects
design by including sector and industry matching directly.

\Cref{tab:dyadic-ladder} reports these six prespecified specifications.
Pooled and firm-effect $R^2$ values are not comparable because the latter
absorb endpoint margins; partial $R^2$ therefore compares the focal regressor
only with the restricted firm-effect model.

\begin{table}[H]
\centering
\small
\begin{tabularx}{\linewidth}{r>{\raggedright\arraybackslash}Xrrrrrr}
\toprule
Spec. & Adjustment & $\hat\beta_{W_2}$ & $\Delta D$/1 SD & 95\% CI & $p$ & $R^2_{\mathrm{full}}$ & $R^2_{\mathrm{partial}}$ \\
\midrule
\multicolumn{8}{l}{\textit{Panel A: staged pooled adjustments}} \\
1 & W2 distance only & 1.746 & 0.064 & [1.302, 2.271] & 0.001 & 0.258 & \textemdash \\
2 & + same-sector indicator & 1.373 & 0.050 & [0.964, 1.853] & 0.001 & 0.288 & \textemdash \\
3 & pooled controls & 1.461 & 0.054 & [1.074, 1.953] & 0.001 & 0.298 & \textemdash \\
4 & + article-count controls & 1.461 & 0.054 & [1.074, 1.953] & 0.001 & 0.298 & \textemdash \\
5 & + within-firm dispersion ($B_i+B_j$, $|B_i-B_j|$) & 1.782 & 0.065 & [1.351, 2.244] & 0.001 & 0.324 & \textemdash \\
\midrule
\multicolumn{8}{l}{\textit{Panel B: canonical endpoint-adjusted specification}} \\
6 & W2 metric + symmetric firm effects & 1.874 & 0.069 & [1.516, 2.250] & 0.001 & 0.742 & 0.332 \\
\bottomrule
\end{tabularx}
\caption{Dyadic association ladder. Every interval and two-sided sign-tail probability uses the same deterministic multinomial node-count array across rungs. Panel A separates the W2 metric's staged pooled adjustments from canonical specification 6, which absorbs symmetric additive firm effects. Specification 5 conditions on the within-firm dispersion components of article geometry and is a decomposition diagnostic. $\Delta D$/1 SD is the fitted change in return chord distance for a one-standard-deviation change in rooted W2 distance. $R^2_{\mathrm{full}}$ is the full-model coefficient of determination; the pooled and fixed-effect blocks absorb different variation and are compared within block. $R^2_{\mathrm{partial}} = 1 - \mathrm{SSE}_{\mathrm{FE}+W_2}/\mathrm{SSE}_{\mathrm{FE}}$ is the conditional fit against the restricted firm-effect model and is undefined (\textemdash) for the pooled rungs, which have no such restricted model.}
\label{tab:dyadic-ladder}
\end{table}

The focal coefficient remains positive throughout, showing that none of the
declared adjustments reverses the sample association.

To distinguish substantive specification changes from rank or conditioning
problems, \Cref{tab:dyadic-design-diagnostics} records the observation count,
design rank, scaled conditioning, and residualized W$_2$ variation for every
specification.
The rank-cleaned canonical design omits one firm indicator, and the
Frisch--Waugh--Lovell coefficient agrees with the direct least-squares fit.

\begin{table}[H]
\centering
\footnotesize
\begin{tabularx}{\linewidth}{@{}lX@{}}
\toprule
\multicolumn{2}{@{}l}{\textit{Panel A: design contract}} \\
\midrule
Sample timing & Pooled news: \ppvalue{sample-year-start}--\ppvalue{sample-year-end}; complete-case returns: \ppvalue{return-panel-n-observations} common dates, \ppvalue{return-panel-date-start}--\ppvalue{return-panel-date-end}. News and returns are pooled over the same window. \\
Observational units & 100 firms and 4950 unordered dyads $i<j$; one observation per pair and no self-dyads. \\
Metrics & $W_{ij}$ is rooted balanced quadratic Wasserstein distance between row-normalized article clouds under Euclidean chord ground cost; $D_{ij}=\sqrt{2(1-\hat\rho_{ij})}$ from complete-case daily returns. \\
Primary design & Specification 6: $D_{ij}=\mu+a_i+a_j+\beta_{W_2} W_{ij}+u_{ij}$, with symmetric additive firm effects and one firm indicator omitted. \\
Resampling & 1999 multinomial node-bootstrap draws; firms are resampled and dyads receive endpoint-multiplicity weight $w_iw_j$, without artificial self-dyads. \\
Estimand & Same-window contemporaneous association conditional on realized $W$ and $D$ and additive firm margins. \\
\bottomrule
\end{tabularx}
\medskip
\begin{tabularx}{\linewidth}{rrrrrX}
\toprule
\multicolumn{6}{l}{\textit{Panel B: pre-fit conditioning by ladder rung}} \\
\midrule
Spec. & Rank/cols. & Scaled $\kappa$ & VIF$_{W_2}$ & FWL SD ratio & Dropped \\
\midrule
1 & 2/2 & 1.00 & 1.00 & 1.000 & \textemdash \\
2 & 3/3 & 1.81 & 1.39 & 0.847 & \textemdash \\
3 & 4/4 & 1.88 & 1.46 & 0.828 & \textemdash \\
4 & 4/4 & 1.88 & 1.46 & 0.828 & \textemdash \\
5 & 6/6 & 2.31 & 1.84 & 0.738 & \textemdash \\
6 & 101/101 & 13.40 & 2.31 & 0.657 & \textemdash \\
\bottomrule
\end{tabularx}
\caption{Dyadic design and pre-fit audit. Panel A fixes the analysis sample, metrics, specification, resampling unit, and interpretation boundary. Panel B reports conditioning over the complete graph on 100 firms: 4950 dyads, degree 99 for every firm, and 0.040 of unordered dyad pairs share an endpoint. $\kappa$ is computed after centring and unit-norm scaling non-intercept columns. VIF$_{W_2}$ and the FWL residual-SD ratio measure how much W2-distance variation remains after the other regressors and summarize numerical conditioning.}
\label{tab:dyadic-design-diagnostics}
\end{table}

With the point design audited, the remaining inference issue is dependence
among dyads that share a firm.
For bootstrap draw $b$, the $N=\ppvalue{n-firms}$ firms receive counts
\begin{equation}\label{eq:node-bootstrap-counts}
  (w_{1b},\ldots,w_{Nb})
  \sim\operatorname{Multinomial}(N;1/N,\ldots,1/N),
\end{equation}
and dyad $(i,j)$ receives frequency weight $w_{ib}w_{jb}$. The two-sided
sign-tail probability is
\begin{equation}\label{eq:dyadic-bootstrap-tail}
  p_{\mathrm{boot}}
  =2\min\left\{
    \frac{1+\sum_{b=1}^{B}\1\{\widehat\beta_{W_2,b}\leq0\}}{B+1},
    \frac{1+\sum_{b=1}^{B}\1\{\widehat\beta_{W_2,b}\geq0\}}{B+1}
  \right\}\wedge1.
\end{equation}
Resampling firms rather than dyads preserves the shared-endpoint dependence:
when firm $i$ appears $w_{ib}$ times, every incident dyad inherits that
multiplicity through $w_{ib}w_{jb}$.

The add-one correction prevents zero simulated tail probabilities
\citep{phipson2010permutation}. Percentile intervals are first-order summaries
of firm-level sampling uncertainty conditional on the constructed article
and return matrices.

\subsection{Representation and article-sampling sensitivity}
\label{sec:representation-diagnostics}
\label{sec:wasserstein-sensitivity}
\label{sec:xmodel-robustness}
\label{sec:geometry-diagnostics}

Having audited the estimator and inference, we next ask whether the reported
association depends on the particular encoder, width, or distance.
\Cref{tab:w2-representation-sensitivity} holds the estimator, return matrix,
firm sample, and node-bootstrap schedule fixed while changing encoder, width,
or distance.
Rank agreement is Spearman correlation with the canonical Qwen3-Embedding-8B,
128-article W$_2$ matrix; the standardized effect is the change in return
chord distance associated with one dyadic standard deviation of the
row's metric.
Writing this effect as $s_r$ for representation $r$, every paired contrast is
$\Delta_{c,r}=s_c-s_r$; a negative value therefore means that the candidate
has the smaller standardized association.
All cell-level association intervals remain positive.
Within these comparisons, W$_1$ nearly preserves the canonical ordering,
while energy produces a materially different ordering but a similar positive
association.

The lower-cost distance comparators reuse the same deterministic article
samples and Euclidean chord ground metric, so their formulas isolate the
distance functional.
Let $A_{ij}$ denote the cross-sample average chord distance and $V_i,V_j$ the
corresponding within-firm averages.
The reported sample $V$-statistic is
\begin{equation}\label{eq:sample-energy-distance}
  \widehat{\mathcal E}_{ij}
  :=\mathcal E(\widehat Q_i,\widehat Q_j) = \sqrt{
    \frac{2}{N_i N_j} \sum_{n=1}^{N_i} \sum_{m=1}^{N_j}
    d_\Omega(\bm{e}_{i,n}, \bm{e}_{j,m}) - V_i - V_j
  }
\end{equation}
where $d_\Omega(\bm{x}, \bm{y}) = \norm{\bm{x} - \bm{y}}_2$ is the Euclidean
ground metric in $\R^{\ppvalue{embed-dim}}$. Each article embedding is
$L^2$-normalized to unit length, so $d_\Omega$ is the chordal metric on the
unit sphere and $V_i,V_j$ measure within-firm coverage dispersion. The
radicand is nonnegative for the biased $V$-statistic, and the resulting matrix
is symmetric with zero diagonal.

The controlled $W_1$ comparator is the equal-weight balanced assignment
\begin{equation}\label{eq:empirical-wasserstein}
  \widehat W_{1,m}(i,j)
  = \min_{\pi\in S_m}\frac{1}{m}\sum_{a=1}^{m}
  \norm{x_{ia}-x_{j,\pi(a)}}_2.
\end{equation}
It is the minimum average semantic displacement needed to align two firms'
article distributions. Unlike W$_2$, it does not give larger displacements
quadratically greater weight.

The close rank agreement between these costs motivates a hypothesis about
representation geometry.
Row normalization maps a nonzero vector $x$ to $x/\norm{x}$ on
$S^{d-1}\subset\mathbb R^d$, with $d=\ppvalue{embed-dim}$ for the canonical
encoder, and bounds chord distances between zero and two.
The Qwen training objective uses cosine similarity
\citep{qwen3_embedding_2025}; more generally, contrastive learning on
normalized representations encourages alignment of related examples and
uniformity across the sphere \citep{wang_isola_hypersphere_2020}.
We conjecture that normalization and the resulting distribution of matching
distances contribute to the observed W$_1$/W$_2$ agreement.
Over a narrow range of positive distances, squared cost is approximately
affine in distance, giving the two assignment objectives similar weights.
Normalization alone neither establishes that concentration nor fixes the
ordering of optimal assignment costs.
The measurement question is whether a representation can retain meaningful
variation in economic magnitude alongside semantic direction.
It motivates representations designed for economic measurement and a
separate assessment of the information carried by vector length.

The seven-row encoder panel targets three measurement concerns.
The capacity comparison asks whether model scale drives the association by
pairing Qwen3-Embedding-8B with its 4B variant at the same native width and
at a common 1,024-dimensional output.
The output-width comparison asks whether the estimated association depends on
the full coordinate set.
It uses Matryoshka prefixes, for which the encoder's training makes
leading-coordinate prefixes usable as nested, lower-capacity embeddings at
256 and 64 coordinates.
These prefixes are fixed before the analysis, not principal components or
selected coordinates \citep{kusupati_matryoshka_2022}.
The architecture comparison asks whether the association is specific to the
Qwen training pipeline.
It uses BGE-large-v1.5, a bidirectional encoder from a different provider and
training pipeline that changes architecture, pretraining, and provider
jointly \citep{xiao_cpack_2023}.
Paired intervals and Holm-adjusted $p$-values compare standardized
associations across representations.
W$_1$ and energy are substituted for $X_{ij}$ only in the descriptive
representation table.
Because all rows share the node schedule, endpoint effects, and omitted
indicator, their standardized effects are comparable even though one-unit
coefficients inherit different distance scales.

\begin{table}[H]
\centering
\footnotesize
\setlength{\tabcolsep}{3.5pt}
\begin{tabular}{lrrrcr}
\toprule
Encoder & Dim. & Distance & Rank $\rho$ & $\Delta D$/1 SD [95\% CI] & Partial $R^2$ \\
\midrule
\multicolumn{6}{l}{\textit{Panel A: encoder and embedding dimension}} \\
Qwen3-8B & 4096 & $W_2$ & 1.000 & 0.069 [0.056, 0.082] & 0.332 \\
Qwen3-4B & 2560 & $W_2$ & 0.965 & 0.068 [0.055, 0.082] & 0.299 \\
Qwen3-8B & 1024 & $W_2$ & 0.985 & 0.067 [0.055, 0.081] & 0.333 \\
Qwen3-4B & 1024 & $W_2$ & 0.960 & 0.066 [0.054, 0.080] & 0.308 \\
BGE-large-v1.5 & 1024 & $W_2$ & 0.904 & 0.064 [0.050, 0.081] & 0.229 \\
Qwen3-8B & 256 & $W_2$ & 0.925 & 0.062 [0.051, 0.075] & 0.322 \\
Qwen3-8B & 64 & $W_2$ & 0.804 & 0.055 [0.044, 0.066] & 0.279 \\
\midrule
\multicolumn{6}{l}{\textit{Panel B: distance family, canonical encoder and dimension}} \\
Qwen3-8B & 4096 & $W_2$ & 1.000 & 0.069 [0.056, 0.082] & 0.332 \\
Qwen3-8B & 4096 & $W_1$ & 0.999 & 0.069 [0.056, 0.083] & 0.332 \\
Qwen3-8B & 4096 & Energy & 0.289 & 0.085 [0.070, 0.102] & 0.356 \\
\midrule
\multicolumn{6}{l}{\textit{Panel C: published annual ChronoBERT sensitivity}} \\
ChronoBERT 2017 & 768 & $W_2$ & 0.743 & 0.036 [0.025, 0.050] & 0.073 \\
ChronoBERT 2018 & 768 & $W_2$ & 0.732 & 0.033 [0.023, 0.047] & 0.063 \\
ChronoBERT 2019 & 768 & $W_2$ & 0.718 & 0.030 [0.020, 0.044] & 0.055 \\
ChronoBERT 2020 & 768 & $W_2$ & 0.726 & 0.033 [0.022, 0.047] & 0.062 \\
ChronoBERT 2021 & 768 & $W_2$ & 0.727 & 0.031 [0.021, 0.046] & 0.057 \\
ChronoBERT 2022 & 768 & $W_2$ & 0.742 & 0.033 [0.022, 0.048] & 0.063 \\
ChronoBERT 2023 & 768 & $W_2$ & 0.756 & 0.036 [0.024, 0.051] & 0.069 \\
ChronoBERT 2024 & 768 & $W_2$ & 0.745 & 0.035 [0.023, 0.050] & 0.065 \\
\bottomrule
\end{tabular}
\caption{Representation sensitivity under the primary estimator. Every row uses symmetric additive firm effects and the same 1,999 node-bootstrap schedule. Rank agreement is Spearman correlation with the canonical Qwen3-Embedding-8B, 128-article rooted-$W_2$ geometry on the common 100-firm, 4,950-dyad panel. The full-width Qwen rows change capacity and native width jointly; the 8B prefixes isolate Matryoshka truncation; BGE-large changes architecture, training pipeline, and provider. The ChronoBERT panel uses published annual cumulative-training checkpoints with a fixed final-layer token-mean inference policy; it is not a matched-compute causal leakage test. Standardized effects use each row's dyadic standard deviation. $W_1$ and energy are lower-cost descriptive comparators.}
\label{tab:w2-representation-sensitivity}
\end{table}
\begingroup
\footnotesize
\setlength{\tabcolsep}{3.5pt}
\begin{longtable}{@{}p{0.35\linewidth}rrrr@{}}
\caption{Base representation paired contrasts under the primary estimator. Estimates are candidate-minus-reference differences in $\Delta D$ per one dyadic standard deviation, using identical draws from the 1,999 node-bootstrap schedule. $p$ denotes the two-sided add-one sign-tail probability; $p_H$ denotes Holm adjustment separately within the seven base and 28 vintage contrasts. Intervals are marginal, not simultaneous.}
\label{tab:w2-base-contrasts} \\
\toprule
Contrast ($c-r$) & Estimate & 95\% CI & $p$ & $p_H$ \\
\midrule
\endfirsthead
\multicolumn{5}{c}{\tablename\ \thetable\ (continued)} \\
\toprule
Contrast ($c-r$) & Estimate & 95\% CI & $p$ & $p_H$ \\
\midrule
\endhead
\midrule
\endfoot
\bottomrule
\endlastfoot
Qwen3-4B full $-$ Qwen3-8B full & -0.001 & [-0.003, 0.001] & 0.388 & 1.000 \\
Qwen3-4B@1024 $-$ Qwen3-8B@1024 & -0.001 & [-0.004, 0.002] & 0.437 & 1.000 \\
Qwen3-8B@1024 $-$ Qwen3-8B full & -0.001 & [-0.003, 0.000] & 0.070 & 0.350 \\
Qwen3-8B@256 $-$ Qwen3-8B@1024 & -0.005 & [-0.008, -0.002] & 0.001 & 0.007 \\
Qwen3-8B@64 $-$ Qwen3-8B@256 & -0.008 & [-0.012, -0.004] & 0.001 & 0.007 \\
BGE-large-v1.5 full $-$ Qwen3-8B@1024 & -0.003 & [-0.009, 0.003] & 0.321 & 1.000 \\
BGE-large-v1.5 full $-$ Qwen3-4B@1024 & -0.002 & [-0.008, 0.004] & 0.470 & 1.000 \\
\end{longtable}
\endgroup
\begingroup
\footnotesize
\setlength{\tabcolsep}{3.5pt}
\begin{longtable}{@{}p{0.35\linewidth}rrrrrr@{}}
\caption{ChronoBERT annual vintage paired contrasts under the primary estimator. Estimates are candidate-minus-reference differences in $\Delta D$ per one dyadic standard deviation, using identical draws from the 1,999 node-bootstrap schedule. $p$ denotes the two-sided add-one sign-tail probability; $p_H$ denotes Holm adjustment separately within the seven base and 28 vintage contrasts. Intervals are marginal, not simultaneous. Eq. requires strict containment of the full interval in the fixed inherited benchmark-calibrated robustness bound shown; this is not a new ChronoBERT preregistration or a matched-compute causal comparison.}
\label{tab:w2-vintage-contrasts} \\
\toprule
Contrast ($c-r$) & Estimate & 95\% CI & $p$ & $p_H$ & Bound & Eq. \\
\midrule
\endfirsthead
\multicolumn{7}{c}{\tablename\ \thetable\ (continued)} \\
\toprule
Contrast ($c-r$) & Estimate & 95\% CI & $p$ & $p_H$ & Bound & Eq. \\
\midrule
\endhead
\midrule
\endfoot
\bottomrule
\endlastfoot
ChronoBERT 2018 $-$ ChronoBERT 2017 & -0.003 & [-0.004, -0.001] & 0.001 & 0.028 & 0.012 & Yes \\
ChronoBERT 2019 $-$ ChronoBERT 2017 & -0.005 & [-0.008, -0.004] & 0.001 & 0.028 & 0.012 & Yes \\
ChronoBERT 2019 $-$ ChronoBERT 2018 & -0.003 & [-0.004, -0.002] & 0.001 & 0.028 & 0.012 & Yes \\
ChronoBERT 2020 $-$ ChronoBERT 2017 & -0.003 & [-0.005, -0.002] & 0.001 & 0.028 & 0.012 & Yes \\
ChronoBERT 2020 $-$ ChronoBERT 2018 & -0.001 & [-0.002, 0.001] & 0.421 & 1.000 & 0.012 & Yes \\
ChronoBERT 2020 $-$ ChronoBERT 2019 & 0.002 & [0.001, 0.004] & 0.008 & 0.136 & 0.012 & Yes \\
ChronoBERT 2021 $-$ ChronoBERT 2017 & -0.004 & [-0.006, -0.003] & 0.001 & 0.028 & 0.012 & Yes \\
ChronoBERT 2021 $-$ ChronoBERT 2018 & -0.002 & [-0.003, 0.000] & 0.028 & 0.364 & 0.012 & Yes \\
ChronoBERT 2021 $-$ ChronoBERT 2019 & 0.001 & [0.000, 0.003] & 0.096 & 0.768 & 0.012 & Yes \\
ChronoBERT 2021 $-$ ChronoBERT 2020 & -0.001 & [-0.002, 0.000] & 0.070 & 0.700 & 0.012 & Yes \\
ChronoBERT 2022 $-$ ChronoBERT 2017 & -0.003 & [-0.006, 0.000] & 0.049 & 0.539 & 0.012 & Yes \\
ChronoBERT 2022 $-$ ChronoBERT 2018 & 0.000 & [-0.003, 0.002] & 0.910 & 1.000 & 0.012 & Yes \\
ChronoBERT 2022 $-$ ChronoBERT 2019 & 0.003 & [0.001, 0.005] & 0.012 & 0.168 & 0.012 & Yes \\
ChronoBERT 2022 $-$ ChronoBERT 2020 & 0.001 & [-0.002, 0.003] & 0.620 & 1.000 & 0.012 & Yes \\
ChronoBERT 2022 $-$ ChronoBERT 2021 & 0.002 & [0.000, 0.004] & 0.096 & 0.768 & 0.012 & Yes \\
ChronoBERT 2023 $-$ ChronoBERT 2017 & 0.000 & [-0.002, 0.002] & 0.695 & 1.000 & 0.012 & Yes \\
ChronoBERT 2023 $-$ ChronoBERT 2018 & 0.002 & [0.001, 0.004] & 0.006 & 0.108 & 0.012 & Yes \\
ChronoBERT 2023 $-$ ChronoBERT 2019 & 0.005 & [0.003, 0.007] & 0.001 & 0.028 & 0.012 & Yes \\
ChronoBERT 2023 $-$ ChronoBERT 2020 & 0.003 & [0.001, 0.005] & 0.001 & 0.028 & 0.012 & Yes \\
ChronoBERT 2023 $-$ ChronoBERT 2021 & 0.004 & [0.003, 0.006] & 0.001 & 0.028 & 0.012 & Yes \\
ChronoBERT 2023 $-$ ChronoBERT 2022 & 0.002 & [0.001, 0.004] & 0.009 & 0.144 & 0.012 & Yes \\
ChronoBERT 2024 $-$ ChronoBERT 2017 & -0.001 & [-0.003, 0.001] & 0.119 & 0.768 & 0.012 & Yes \\
ChronoBERT 2024 $-$ ChronoBERT 2018 & 0.001 & [0.000, 0.003] & 0.076 & 0.700 & 0.012 & Yes \\
ChronoBERT 2024 $-$ ChronoBERT 2019 & 0.004 & [0.003, 0.006] & 0.001 & 0.028 & 0.012 & Yes \\
ChronoBERT 2024 $-$ ChronoBERT 2020 & 0.002 & [0.001, 0.004] & 0.010 & 0.150 & 0.012 & Yes \\
ChronoBERT 2024 $-$ ChronoBERT 2021 & 0.003 & [0.002, 0.005] & 0.001 & 0.028 & 0.012 & Yes \\
ChronoBERT 2024 $-$ ChronoBERT 2022 & 0.001 & [0.000, 0.004] & 0.163 & 0.815 & 0.012 & Yes \\
ChronoBERT 2024 $-$ ChronoBERT 2023 & -0.001 & [-0.002, 0.000] & 0.034 & 0.408 & 0.012 & Yes \\
\end{longtable}
\endgroup

Moderate truncation from full width to 1,024 coordinates leaves the 8B
standardized association statistically indistinguishable from the canonical
cell. The subsequent 1,024-to-256 and 256-to-64 compression steps are negative
and remain distinguishable from zero after Holm adjustment. Neither capacity
comparison nor either BGE contrast is distinguishable after the same
base-family adjustment. This pattern locates the compression levels at which
the measured
association weakens.

The supplementary lexical comparisons connect the representation checks to
\citet{box2018qualitative} using the same articles and return window.
We compare cosine distance between firms' aggregate TF--IDF vectors with
W$_2$ between equally weighted distributions of TF--IDF article vectors.
Each construction estimates its vocabulary and term weights separately,
with \ppnum[0]{text-controls-max-features} features to match the canonical
Qwen output width, providing a lexical benchmark.

Four further comparisons use the common 610-million-parameter architecture
of \citet{gisserot2026should}.
One model is pretrained to predict the next token (CLM), while Masked Language Modelling (MLM) models predict
masked tokens at masking rates of 30\%, 40\%, and 50\%.
All four use bidirectional attention when representing articles, average
token representations into an article vector, and normalize its length
before computing W$_2$.
Crucially, unlike our other QWEN models, these models receive no contrastive fine-tuning for document similarity, demonstrating the relative efficiency and efficacy of various self-supervised learning techniques, when data, initialization and compute are held fixed. These results replicate the findings of \citet{gisserot2026should} on market-priced tasks, showing that low masking ratios demonstrate the strongest learned representations without further fine-tuning.

\begingroup
\small
\begin{longtable}{lrrr}
\caption{Supplementary lexical and pretraining comparisons on 2018-2022 returns, using 100 firms and 4,950 dyads. All rows use symmetric additive firm effects and the same 1,999 multinomial node-bootstrap draws. $\Delta D$/1 SD is the fitted change in return chord distance per full-sample dyadic standard deviation of each distance, with its pointwise percentile 95\% interval. Partial $R^2$ measures the reduction in residual sum of squares relative to firm effects alone. Rank agreement is Spearman correlation with canonical Qwen $W_2$. These supplementary comparisons are separate from the paired-contrast families reported above.}
\label{tab:text-controls} \\
\toprule
Representation & $\Delta D$/1 SD [95\% CI] & Partial $R^2$ & Rank agreement \\
\midrule
\endfirsthead
\multicolumn{4}{c}{\tablename\ \thetable\ (continued)} \\
\toprule
Representation & $\Delta D$/1 SD [95\% CI] & Partial $R^2$ & Rank agreement \\
\midrule
\endhead
\midrule
\endfoot
\bottomrule
\endlastfoot
\multicolumn{4}{l}{\textit{Panel A: lexical benchmarks and canonical reference}} \\
Qwen3-8B $W_2$ (reference) & 0.069 [0.056, 0.082] & 0.332 & 1.000 \\
TF--IDF firm cosine distance & 0.119 [0.077, 0.181] & 0.100 & 0.099 \\
TF--IDF document $W_2$ & 0.054 [0.040, 0.070] & 0.133 & 0.802 \\
\midrule
\multicolumn{4}{l}{\textit{Panel B: pretraining checkpoints}} \\
610M CLM $W_2$ & 0.037 [0.025, 0.052] & 0.082 & 0.767 \\
610M MLM30 $W_2$ & 0.048 [0.035, 0.064] & 0.106 & 0.764 \\
610M MLM40 $W_2$ & 0.042 [0.030, 0.056] & 0.079 & 0.768 \\
610M MLM50 $W_2$ & 0.023 [0.013, 0.038] & 0.033 & 0.399 \\
\end{longtable}
\endgroup

All supplementary slopes are positive, with pointwise node-bootstrap
intervals excluding zero (\Cref{tab:text-controls}).
The TF--IDF W$_2$ result shows that the association survives simple lexical
representations of article distributions.
Its higher fit than aggregate cosine distance reflects a joint change in
aggregation, weighting, and distance construction.

Within the masked-prediction models, both the standardized association and
partial $R^2$ decline as the masking rate rises from 30\% to 50\%.
The 30\% model has the strongest point-estimate fit among the four base
models; CLM and the 40\% model have similar fit, and the 50\% model has the weakest.
Thus the news--comovement association appears under both training objectives,
with no uniform point-estimate advantage for masked prediction.
More extensive masking does not yield a stronger measured connection to
return dependence in this sample.
These rankings are descriptive; the pointwise intervals test each association,
not differences between objectives or masking rates.
All four base models have lower fit than lexical W$_2$ and canonical Qwen.
These supplementary comparisons use 2018--2022 returns and remain separate
from the paired-contrast families above.

Training-information cutoff requires a separate design because comparisons
across off-the-shelf encoders also change architecture and training pipeline.
We use the eight published ChronoBERT checkpoints dated at each year-end from
2017 through 2024 \citep{he_chronobert_2025}.
These ModernBERT-based encoders were pretrained by masked token prediction.
Later checkpoints continue training the preceding checkpoint on newly
available text; the panel therefore combines later information with cumulative
training exposure, rather than holding token budget or compute fixed.
For each fixed checkpoint, we encode the same 128 articles per firm and average
the final-layer token states over all non-padding positions, including special
tokens, to obtain a 768-coordinate document representation.
The common inference policy retains at most 8,000 characters and 1,024 tokens
per article and uses bfloat16 autocast with float32 model parameters and
token-mean pooling.
We apply row-L2 normalization once before distance construction, without
selecting layers using returns.

The 2017 training cutoff precedes the
\ppvalue{sample-year-start}--\ppvalue{sample-year-end} article window;
2018--2022 cutoffs fall within construction, and 2023--2024 cutoffs follow it.
The last two nevertheless overlap, rather than postdate, the full 2023--2026
return-evaluation period.
Publication occurred later: the comparison concerns reported training
information, not historical deployment availability or point-in-time validity
of every other input.
All 28 later-minus-earlier contrasts use the common node-bootstrap draws, with
Holm adjustment within this vintage family.
Equivalence requires the full marginal 95\% paired interval to lie strictly
inside the symmetric inherited benchmark-calibrated margin reported in
\Cref{tab:w2-vintage-contrasts}.
This is a fixed robustness margin, not a new ChronoBERT preregistration;
the intervals are not simultaneous confidence intervals.
All annual pairs meet this inherited equivalence rule, although some paired
differences reject zero after Holm adjustment.
The estimates do not change monotonically with the cutoff year.
This combination supports bounded sensitivity under the stated margin,
not exact representation invariance or the absence of training leakage.

The Matryoshka diagnostics show how truncation changes the underlying
dyad geometry.
In \Cref{fig:matryoshka-preservation}, the diagonal is exact distance
preservation and $\rho$ measures dyad-rank agreement with full width.
The 256-coordinate representation is closer to the full matrix than the
64-coordinate representation.

\begin{figure}[H]
  \centering
  \import{images/}{matryoshka_distance_preservation\ppfiguresuffix.pgf}
  \caption{Truncated versus full-width Qwen3-Embedding-8B rooted W$_2$ over the
    common firm pairs. Both axes are rooted W$_2$ distance in unit-chord
    embedding space; dashed lines mark equality and $\rho$ is Spearman rank
    agreement with the full-width matrix. The 256-coordinate prefix preserves
  the canonical geometry more closely than the 64-coordinate prefix.}
  \label{fig:matryoshka-preservation}
\end{figure}

Article-sampling checks vary the number of observations representing each firm.
\Cref{tab:w2-sampling-diagnostics} therefore varies the deterministic
per-firm article prefix at fixed encoder and distance.
Standardized effects remain positive and rank agreement increases toward the
128-article benchmark.

\begin{table}[H]
\centering
\footnotesize
\setlength{\tabcolsep}{3.5pt}
\begin{tabular}{lrrrcr}
\toprule
Encoder & Articles & Distance & Rank $\rho$ & $\Delta D$/1 SD [95\% CI] & Partial $R^2$ \\
\midrule
Qwen3-8B & 32 & $W_2$ & 0.895 & 0.060 [0.047, 0.073] & 0.260 \\
Qwen3-8B & 64 & $W_2$ & 0.960 & 0.066 [0.054, 0.079] & 0.308 \\
Qwen3-8B & 128 & $W_2$ & 1.000 & 0.069 [0.056, 0.082] & 0.332 \\
\bottomrule
\end{tabular}
\caption{Article-sampling sensitivity for Qwen3-Embedding-8B. Prefixes are deterministic, rank agreement is measured against the 128-article $W_2$ matrix, and all association intervals reuse the common node schedule. The rows diagnose finite article sampling; they do not quantify embedding-estimation uncertainty.}
\label{tab:w2-sampling-diagnostics}
\end{table}

\subsection{Sector structure and projection validity}
\label{sec:projection-validity}
\label{sec:geometric-representation}
\label{sec:geometric-diagnostics}

Metric multidimensional scaling (MDS) constructs a descriptive two-dimensional
map by minimizing squared discrepancies between projected Euclidean distances
and the original W$_2$ distances between firms.
\Cref{fig:sector-metric-mds} displays the fitted map, whose axes have no assigned
economic meaning.
Its compression can distort both local neighbours and global distances, so
economic pair comparisons and all estimation use the original W$_2$ matrix.

\begin{figure}[htbp]
  \centering
  \import{images/}{metric_mds_sector_plot\ppfiguresuffix.pgf}
  \caption{Descriptive two-dimensional metric MDS of the exact 128-article W$_2$
    matrix for the canonical Qwen3-Embedding-8B representation.
    Both axes are unitless fitted coordinates; each labelled point is a
    priced firm, and colours identify sectors.
    Pair selection and estimation use the original W$_2$ matrix;
    \Cref{fig:metric-mds-shepard-diagnostic} reports projection fidelity.
  }
  \label{fig:sector-metric-mds}
\end{figure}

Sector structure is assessed directly in the canonical W$_2$ geometry by a
one-way permutational multivariate analysis of variance (PERMANOVA)
\citep{anderson_permanova_2001}.
Let $d_{ij}=W_2(Q_i,Q_j)$ for the $N$ observed firms, grouped into $K$ sectors
with sizes $n_g$.
The distance-based decomposition uses squared original distances:
\[
  SS_T=\frac{1}{N}\sum_{i<j}d_{ij}^2,\qquad
  SS_W=\sum_g\frac{1}{n_g}\sum_{\substack{i<j\\i,j\in g}}d_{ij}^2,
  \qquad R^2=1-\frac{SS_W}{SS_T}.
\]
The observed sector share is $R^2=\ppnum[3]{w2-permanova-r2}$, compared with
the fixed-size random-label expectation $(K-1)/(N-1)
=\ppnum[3]{w2-permanova-null-r2}$.
The pseudo-$F$ ratio divides $(SS_T-SS_W)/(K-1)$ by $SS_W/(N-K)$;
its permutation reference reassigns whole firms' labels while preserving sector
sizes and holding the distance matrix fixed.
This post-hoc diagnostic describes sector association conditional on the observed
firms and distances; inference requires exchangeability under the reference null
and does not identify a causal sector effect.

\begin{table}[H]
\centering
\small
\textit{Panel A: original-distance sector decomposition}
\par\smallskip
\begin{tabular}{rrrrrr}
\toprule
$SS_W$ & $SS_B$ & $SS_T$ & $R^2$ & $F_{\rm pseudo}$ & $p_{\rm perm}$ \\
\midrule
46.831 & 7.402 & 54.233 & 0.136 & 1.798 & 0.0001 \\
\bottomrule
\end{tabular}
\par\medskip
\textit{Panel B: random-label reference}
\par\smallskip
\begin{tabular}{rrr}
\toprule
$\mathbb{E}_{\rm perm}[R^2]$ & Mean permuted $R^2$ & Excess $R^2$ \\
\midrule
0.081 & 0.081 & 0.056 \\
\bottomrule
\end{tabular}
\par\medskip
\textit{Panel C: dispersion companion}
\par\smallskip
\begin{tabular}{rrrr}
\toprule
Firms & Sectors & $F_{\rm disp}$ & $p_{\rm disp}$ \\
\midrule
98 & 7 & 11.494 & 0.0001 \\
\bottomrule
\end{tabular}
\caption{One-way PERMANOVA on the canonical rooted W$_2$ matrix for 100 firms in 9 sectors. Sums of squares use squared original W$_2$ distances, without projecting or estimating Wasserstein barycentres. The reference distribution uses 9999 whole-firm label permutations preserving sector sizes, and add-one upper-tail $p$-values. The analytic random-label expectation is $(K-1)/(N-1)$; excess $R^2$ subtracts the simulated null mean. These are descriptive null references, not confidence intervals or causal sector effects; permutation inference requires exchangeability under the reference null. The spread companion retains 98 firms in 7 replicated sectors; excluded singleton sectors are Energy, Financial Services. It uses all positive principal-coordinate axes, within-sector spatial medians, the $\sqrt{n_g/(n_g-1)}$ radius adjustment, and 9999 residual permutations of the scalar sector model. A large dispersion $p$-value does not establish equal sector spreads. Sector sizes are Communication Services=11, Consumer Cyclical=18, Consumer Defensive=7, Energy=1, Financial Services=1, Healthcare=14, Industrials=7, Technology=38, Utilities=3.}
\label{tab:w2-permanova}
\end{table}

The dispersion companion in \Cref{tab:w2-permanova} compares distances to
within-sector spatial medians in the full principal-coordinate representation
\citep{anderson_dispersion_2006}.
Singleton sectors remain in the main decomposition but cannot estimate
within-sector spread and are excluded from this companion.
Unequal sector sizes make spread differences consequential for interpreting the
grouping test; a nonsignificant companion does not establish equal spreads.
The grouping test has $p=\ppnum[4]{w2-permanova-p}$, and the dispersion
companion has $p=\ppnum[4]{w2-permanova-dispersion-p}$.
The detected sector organization may therefore include spread differences, so
the main test does not isolate sector location.
\Cref{fig:metric-mds-shepard-diagnostic} separately assesses whether the
two-dimensional display preserves the original distances and neighbours.

\begin{figure}[H]
  \centering
  \import{images/}{metric_mds_shepard_diagnostic\ppfiguresuffix.pgf}
  \caption{Fidelity of the canonical two-dimensional W$_2$ metric-MDS map. The
    horizontal axis is exact pairwise W$_2$ and the vertical axis is projected
    Euclidean distance $\hat d_{ij}$. Stress-1,
    $\{\sum_{i<j}(d_{ij}-\hat d_{ij})^2/\sum_{i<j}d_{ij}^2\}^{1/2}$,
    is \ppnum[3]{metric-mds-stress-1}; Pearson and
    Spearman Shepard correlations are \ppnum[3]{metric-mds-shepard-pearson} and
    \ppnum[3]{metric-mds-shepard-spearman}. Mean top-five-neighbour recall is
    \ppnum[3]{metric-mds-top5-neighbour-recall}, and the exact nearest neighbour is
    preserved for \ppvalue{metric-mds-exact-nearest-preserved} of
    \ppvalue{n-firms} firms. Pair selection uses the original distance matrix.
  }
  \label{fig:metric-mds-shepard-diagnostic}
\end{figure}

Principal coordinates analysis (PCoA) provides a separate spectral diagnostic.
It double-centres the squared W$_2$ matrix, multiplies by $-1/2$, and
eigendecomposes the result.
The full spectrum distinguishes concentration in a few positive axes from
geometry that a Euclidean map cannot reproduce.
For the canonical representation, its first two positive axes retain
\ppnum[3]{pcoa-two-axis-share} of total positive inertia; this share does not
describe the metric-MDS coordinates.
For each representation, an axis's positive-inertia share divides its positive
eigenvalue by the sum of all positive eigenvalues, including axes beyond those
shown in the scree panel.
Material negative eigenvalues indicate non-Euclidean structure in the squared
distance matrix; their absolute sum divided by the sum of absolute eigenvalues
measures its negative-inertia share.
For the canonical representation this share is
\ppnum[3]{pcoa-negative-inertia-share}.
It is reported separately rather than folded into explained positive inertia.

\Cref{fig:pcoa-scree,fig:pcoa-vintage-scree} compare spectral concentration
across the seven encoder/width cells and the eight annual ChronoBERT cells,
respectively, using the same firms and each cell's 128-article W$_2$ matrix.
The return benchmark uses the eigenvalues of the same firms' unstandardized
in-sample covariance matrix.
This is PCA of returns with dates as observations and firms as variables.
Its eigenvalue shares measure explained return variance, whereas the W$_2$
curves measure positive inertia in firm-distance geometry.
Their comparison describes spectral concentration and does not measure return
variance explained by news coordinates.
\Cref{tab:pcoa-spectrum} reports the two-axis positive-inertia and negative-mass
shares for all fifteen representations.

\begin{figure}[H]
  \centering
  \import{images/}{pcoa_scree\ppfiguresuffix.pgf}
  \caption{Spectral concentration of the seven base encoder/width W$_2$
    geometries and in-sample return PCA on the common firm universe.
    The upper panel overlays the first twenty positive-axis shares; the lower
    panel overlays cumulative shares over the full available spectra.
    Each W$_2$ eigenvalue is divided by its matrix's total positive inertia;
    return PCA eigenvalues are divided by total return variance.
    The seven geometries are Qwen3-8B at 4,096, 1,024, 256 and 64 coordinates,
    Qwen3-4B at 2,560 and 1,024 coordinates, and BGE-large-v1.5 at 1,024
    coordinates, each using 128 articles per firm.
    Return PCA uses the unstandardized in-sample covariance matrix.
    Negative W$_2$ inertia is reported separately in
    \Cref{tab:pcoa-spectrum}; the curves do not estimate return variance
  explained by news. }
  \label{fig:pcoa-scree}
\end{figure}

\begin{figure}[H]
  \centering
  \import{images/}{pcoa_vintage_scree\ppfiguresuffix.pgf}
  \caption{Spectral concentration of the eight ChronoBERT W$_2$ geometries
    and the same in-sample return-PCA benchmark as \Cref{fig:pcoa-scree}.
    Each annual checkpoint from 2017 through 2024 represents the same 128
    articles per firm in 768 coordinates.
    The upper panel overlays the first twenty positive-axis shares and the
    lower panel cumulative shares over the full available spectra.
    W$_2$ shares use total positive inertia as their denominator; return-PCA
    shares use total variance of the unstandardized in-sample covariance
    matrix. Negative W$_2$ inertia is reported separately in
    \Cref{tab:pcoa-spectrum}.
    These cumulative-training checkpoints do not hold training exposure or
  compute fixed. }
  \label{fig:pcoa-vintage-scree}
\end{figure}

\begin{table}[H]
\centering\small
\caption{Spectral concentration of the Paper 1 news geometries. The first two axes are a share of total positive inertia; negative inertia is a share of absolute spectral mass.}
\label{tab:pcoa-spectrum}
\begin{tabular}{lrrr}
\toprule
Model / vintage & Width & First two axes (\%) & Negative mass (\%) \\
\midrule
Qwen3-8B & 4,096 & 6.4 & 0.0 \\
Qwen3-4B & 2,560 & 6.2 & 0.0 \\
Qwen3-8B & 1,024 & 6.5 & 0.0 \\
Qwen3-4B & 1,024 & 6.6 & 0.0 \\
BGE-large-v1.5 & 1,024 & 6.3 & 0.0 \\
Qwen3-8B & 256 & 7.1 & 0.0 \\
Qwen3-8B & 64 & 10.1 & 0.0 \\
ChronoBERT 2017 & 768 & 12.9 & 0.0 \\
ChronoBERT 2018 & 768 & 13.4 & 0.0 \\
ChronoBERT 2019 & 768 & 13.8 & 0.0 \\
ChronoBERT 2020 & 768 & 13.0 & 0.0 \\
ChronoBERT 2021 & 768 & 13.4 & 0.0 \\
ChronoBERT 2022 & 768 & 13.8 & 0.0 \\
ChronoBERT 2023 & 768 & 13.1 & 0.0 \\
ChronoBERT 2024 & 768 & 13.3 & 0.0 \\
\bottomrule
\end{tabular}
\end{table}

\subsection{Returns of neighbouring firms}
\label{sec:exploratory-geometry-outcomes}

\Cref{fig:neighbour-outcome-paths} traces cumulative log returns for
\ppnum[0]{neighbour-holdout-n-selected} within-sector pairs selected as
nearest neighbours in the 2018--2020 W$_2$ article geometry.
Some pairs track closely over 2021--2022, while others separate materially.
The display illustrates the variety of realized dependence among firms with
similar reported information.

\begin{figure}[H]
  \centering
  \import{images/}{neighbour_holdout_paths\ppfiguresuffix.pgf}
  \caption{Cumulative log returns for
    \ppnum[0]{neighbour-holdout-n-selected} within-sector nearest-neighbour
    pairs selected using 2018--2020 article W$_2$.
    Panels label sectors and tickers over 4 January 2021--30 December 2022.
    Insets show W$_2$, RMS and terminal path gaps, and RMS percentile among
    within-sector pairs (zero denotes closest tracking).
  }
  \label{fig:neighbour-outcome-paths}
\end{figure}

\section{Transport interpretations and finite-support
transfer}\label{app:finite-support-transfer}

\subsection{Valuations and adjustment costs}
\label{sec:transport-economic-interpretation}

For the economic interpretations, consider probability laws $\mu,\nu$ with
compact support in Euclidean space and a continuous nonnegative transport
cost $c(x,y)$.
Write $\Pi(\mu,\nu)$ for their couplings.
Kantorovich duality equates the least allocation cost with the greatest
supporting valuation \citep{panaretos_invitation_2020}:
\begin{equation}\label{eq:transport-economic-dual}
  \inf_{\pi\in\Pi(\mu,\nu)}\int c(x,y)\,d\pi
  =\sup_{\varphi(x)+\psi(y)\le c(x,y)}
  \left\{\int\varphi\,d\mu+\int\psi\,d\nu\right\}.
\end{equation}
The supremum ranges over continuous origin and destination potentials on
the compact supports.
The inequality makes their combined valuation a lower bound on every
feasible shipment cost; optimizing these valuations attains the primal value.
For metric cost $c(x,y)=\norm{x-y}$, the Kantorovich--Rubinstein
specialization uses a single valuation function:
\begin{equation}\label{eq:transport-economic-w1}
  W_1(\mu,\nu)
  =\sup_{|f(x)-f(y)|\le\norm{x-y}}
  \left\{\int f\,d\mu-\int f\,d\nu\right\}.
\end{equation}
The constraint is the spatial-arbitrage bound for symmetric unit transport
costs.
Quadratic cost uses the two-potential dual in
\eqref{eq:transport-economic-dual}.
These potentials value transport constraints; $L$ in the main text bounds
the separate transmission of characteristics into exposures.

The dynamic interpretation of quadratic cost considers paths of probability
laws $\mu_t$ and velocity fields $v_t$ in the ambient Euclidean space.
The Benamou--Brenier formula is
\begin{equation}\label{eq:transport-economic-action}
  W_2^2(\mu,\nu)
  =\inf_{\mu_t,v_t}\int_0^1\int\norm{v_t(x)}^2\,d\mu_t(x)\,dt,
  \qquad
  \partial_t\mu_t+\nabla\!\cdot(v_t\mu_t)=0,
\end{equation}
with endpoints $\mu_0=\mu$ and $\mu_1=\nu$, and the continuity equation
understood weakly \citep{benamou_brenier_2000}.
An instantaneous cost $\norm{v}^2/2$ therefore gives minimum cost $W_2^2/2$.
Unit-normalized endpoints retain this interpretation with paths in ambient
space, consistent with the empirical chord metric.

To make the curvature connection precise, let an adjustment penalty $a(v)$
be twice continuously differentiable near its minimum at $v=0$, with
positive-definite Hessian $Q$.
Then $a(v)-a(0)=\tfrac12 v^\top Qv+o(\norm{v}^2)$.
The change of coordinates $z=Q^{1/2}x$ expresses the local quadratic
penalty in Euclidean units.
This supplies the adjustment-cost interpretation; the paper's exposure
geometry is fixed separately by the common-shock covariance operator $\Gamma$.

\subsection{Transfer for finite characteristic clouds}

The metric bracket in \Cref{cor:random-w2-metric} supplies the main-text
covariance restriction.
The following complementary bound compares squared transport costs directly
for indexed finite clouds with matching atom masses and a bounded support diameter.

\begin{proposition}[Finite-support transfer]\label{prop:random-w2-transfer}
  Write each finite characteristic cloud as
  \[
    \widehat C_i=\sum_a p_{ia}\delta_{x_{ia}},
    \qquad
    \widehat C_j=\sum_b p_{jb}\delta_{x_{jb}}.
  \]
  Fix one transmission draw $u_{ia}$ for each characteristic atom and define
  the indexed exposure clouds by
  \[
    \widehat P_i:=\sum_a p_{ia}\delta_{T_i(x_{ia},u_{ia})},
    \qquad
    \widehat P_j:=\sum_b p_{jb}\delta_{T_j(x_{jb},u_{jb})}.
  \]
  $G_{ij}:=W_{2,d_\Omega}(\widehat C_i,\widehat C_j)$ and
  $\tau_{ij}:=\tau_i+\tau_j$. Let $D_{ij}$ be any maintained bound on the
  risk-coordinate support diameter, that is on
  $\norm{t_\Gamma(x)-t_\Gamma(y)}$ and on $\norm{Z_i-Z_j}$ across all pairs of
  indexed support points. Then, under \Cref{ass:random-common-map},
  \begin{align}
    L^{-2}G_{ij}^{2}-2\tau_{ij}D_{ij}-\tau_{ij}^{2}
    &\le W_{2,\Gamma}^{2}(\widehat P_i,\widehat P_j)
    \label{eq:random-w2-lower-transfer}\\
    W_{2,\Gamma}^{2}(\widehat P_i,\widehat P_j)
    &\le L^{2}G_{ij}^{2}+2\tau_{ij}D_{ij}+\tau_{ij}^{2}.
    \label{eq:random-w2-upper-transfer}
  \end{align}
\end{proposition}

\begin{proof}
  Let $M=(m_{ab})$ be a transport matrix. Because the indexed characteristic
  and exposure clouds use the same atom masses, every nonnegative matrix with
  row sums $p_{ia}$ and column sums $p_{jb}$ is feasible for both the
  characteristic and exposure problems. Thus both Wasserstein costs minimize
  over the same coupling-matrix polytope.
  Write $Y_{ia}:=t_\Gamma(x_{ia})$ and
  $Z_{ia}:=\Gamma^{1/2}T_i(x_{ia},u_{ia})$, and define $Y_{jb},Z_{jb}$
  analogously. The indexed slack bounds imply
  $\norm{Z_{ia}-Y_{ia}}\le\tau_i$ and
  $\norm{Z_{jb}-Y_{jb}}\le\tau_j$.
  For each entry $(a,b)$, the bi-Lipschitz inequality gives
  \[
    L^{-2}d_\Omega(x_{ia},x_{jb})^2
    \le \norm{t_\Gamma(x_{ia})-t_\Gamma(x_{jb})}^2
    \le L^2d_\Omega(x_{ia},x_{jb})^2.
  \]
  The indexed slack vectors have norms at most $\tau_i$ and $\tau_j$, so the
  triangle inequality gives, entry by entry,
  \[
    \left|
    \norm{Z_{ia}-Z_{jb}}^2
    -\norm{t_\Gamma(x_{ia})-t_\Gamma(x_{jb})}^2
    \right|
    \le 2\tau_{ij}D_{ij}+\tau_{ij}^2.
  \]
  Combining the two entrywise displays, average against an arbitrary feasible
  $M$. Because every admissible $M$ is available in both problems, minimizing
  the lower and upper comparisons over the shared polytope gives
  equations~\ref{eq:random-w2-lower-transfer}--\ref{eq:random-w2-upper-transfer}.
\end{proof}

The finite-support proposition shows why matching atom masses matter: the
same transport weights can be evaluated before and after transmission, so
bounded distortion and slack carry characteristic-cloud distance into an
exposure-distance interval.
Economically, the interval is conditional on the declared values of $L$,
$\tau_{ij}$, and $D_{ij}$; permitting more transmission distortion or a
larger support bound weakens what a finite characteristic cloud can imply
about its indexed exposure cloud.

\clearpage
\section{Formal verification}\label{sec:lean-proofs}

Analytical proofs accompany the covariance endpoints and metric bracket in
\Cref{sec:methodology}, and the finite-support bounds in
\Cref{app:finite-support-transfer}.
The optimal-transport existence result under the stated Polish-space and
second-moment conditions supplies the optimizer for
\Cref{thm:random-exposure-envelope} \citep[Ch.~1]{panaretos_invitation_2020}.
The formalization takes an optimal-plan certificate as an input.
The machine-checked statements and their scope are recorded below.

\ppLeanStatusList

\end{document}